\documentclass[10pt,twocolumn,twoside]{IEEEtran} 

\usepackage{cite}
\usepackage{amsmath,amssymb,amsfonts}
\usepackage{algorithmic}
\usepackage{graphicx}
\usepackage{textcomp}
\usepackage{xcolor}
\usepackage{epstopdf}
\usepackage{import}
\usepackage{booktabs}
\usepackage{xspace}
\usepackage{multirow}
\usepackage[normalem]{ulem}
\usepackage[absolute,showboxes]{textpos}
\usepackage{tikz}

\newcommand{\MAT}{\textsc{Matlab}}

\ifCLASSINFOpdf
\else
\fi
\ifCLASSOPTIONcompsoc
\usepackage[caption=false,font=normalsize,labelfont=sf,textfont=sf]{subfig}
\else
\usepackage[caption=false,font=footnotesize]{subfig}
\fi
\TPMargin{3pt}
\newcommand{\copyrightstatementa}{
	\begin{textblock}{0.65}(0.18,0.016)    
		\noindent
		\footnotesize
		\textbf{Please cite this version:} \\
		M. Kumru and E. Özkan, ``Three-Dimensional Extended Object Tracking and Shape Learning Using Gaussian Processes," IEEE Transactions on Aerospace and Electronics, 2021. DOI: 10.1109/TAES.2021.3067668.
	\end{textblock}
}
\newcommand{\copyrightstatementb}{
	\begin{textblock}{0.84}(0.08,0.954)    
		\noindent
		\footnotesize
		\copyright 2021 IEEE. Personal use of this material is permitted. Permission from IEEE must be obtained for all other uses, in any current or future media, including reprinting/republishing this material for advertising or promotional purposes, creating new collective works, for resale or redistribution to servers or lists, or reuse of any copyrighted component of this work in other works. 
	\end{textblock}
}

\begin{document}
	
	\copyrightstatementa
	\copyrightstatementb
	%
	\title{Three-Dimensional Extended Object Tracking and Shape Learning Using Gaussian Processes}
	%
	%
	%
	
	\author{Murat Kumru
		and Emre \"{O}zkan,~\IEEEmembership{Member,~IEEE}
		\thanks{The authors are with the Department
			of Electrical and Electronics Engineering, Middle East Technical University, 06531, Ankara,
			Turkey (e-mail: kumru@metu.edu.tr; emreo@metu.edu.tr).}
	}
	
	%
	%

	\markboth{}%
	{Kumru and  \"{O}zkan: Three-Dimensional Extended Object Tracking and Shape Learning Using Gaussian Processes}
	%



	\maketitle
	
	\begin{abstract}
		In this study, we investigate the problem of tracking objects with unknown shapes using three-dimensional (3D) point cloud data. 
		We propose a Gaussian process-based model to jointly estimate object kinematics, including position, orientation and velocities, together with the shape of the object for online and offline applications. 
		We describe the unknown shape by a radial function in 3D, and induce a correlation structure via a Gaussian process. 
		Furthermore, we propose an efficient algorithm to reduce the computational complexity of working with 3D data. 
		This is accomplished by casting the tracking problem into projection planes which are attached to the object's local frame. 
		The resulting algorithms can process 3D point cloud data and accomplish tracking of a dynamic object. 
		Furthermore, they provide analytical expressions for the representation of the object shape in 3D, together with confidence intervals.
		The confidence intervals, which quantify the uncertainty in the shape estimate, can later be used for solving the gating and association problems inherent in object tracking. 
		The performance of the methods is demonstrated both on simulated and real data. 
		The results are compared with an existing random matrix model, which is commonly used for extended object tracking in the literature. 	
	\end{abstract}
	
	\begin{IEEEkeywords}
		Extended Object Tracking, Gaussian Processes, Shape Learning, Point Cloud Data.
	\end{IEEEkeywords}
	
	\IEEEpeerreviewmaketitle
	
	\section{Introduction}
    Object tracking can be described as making inference about the unknown kinematic properties of an object using sequentially available sensor data. 
    The problem is explicitly referred to as \textit{point object tracking} when the object of interest is relatively small that it may occupy a single cell of the sensor device. 
    In this case, the number of measurements returned from the object is limited to be at most one per scan. 
    On the other hand, it is dubbed \textit{extended object tracking} (EOT) when the resolution of the sensor is sufficiently high relative to the size of the object so that it may be detected by several cells of the sensor. 
    Thus, the object potentially originates multiple measurements at a single scan in this setting. 
    	
    There is a solid body of literature considering the EOT problem, specifically in two-dimensional (2D) space (see \cite{granstrom2016extended} for a comprehensive survey). 
    The methods of this category process 2D measurements by relying on various extent models. 
    These extent representations exhibit significant variance in their compactness and expressive power. 
    For example, a group of EOT algorithms imposes simple geometric shape models such as a circle, a rectangle, or an ellipse, \cite{petrov2011novel, petrovskaya2009model, koch2008bayesian, feldmann2011tracking, orguner2012variational, lan2012tracking, granstrom2015extended}. 
    These essentially achieve extent modeling with only a few parameters at the cost of limited potential for shape description. 
    In another line of research, random hyper-surface models (RHM) suggest a more flexible extent representation for star-convex objects, \cite{baum2009random}, \cite{baum2011shape}. 
    The model is based on the Fourier series expansion of the spatial extent, and the coefficients of the expansion are estimated together with the kinematics. 
    Similarly, the latent extent of a star-convex object is described by a Gaussian process (GP) in \cite{wahlstrom2015extended} and \cite{ozkan2016rao}. 
    With its favorable analytical properties and close connections to the Bayesian paradigm, this model enables effective estimation of the object pose and its shape. 
    Several adaptations of this approach have been investigated for various application settings, \cite{hirscher2016multiple, guo2020simultaneous, thormann2018extended, tuncer2018extended, tuncer2019extended, lee2019extended, michaelis2017heterogeneous, michaelis2019merging}. 
    
    On the other hand, there have been remarkably few attempts to tackle the EOT problem in three-dimensional (3D) space, despite the increasing accessibility of sensors such as depth cameras and LIDARs generating 3D data. 
    Furthermore, the majority of these attempts are direct generalizations of the basic geometric models and represent the extent via, for example, an ellipsoid or a bounding box. 
    One of the few exceptions that aim at a more generic extent description is proposed in \cite{faion2015recursive}. 
    The method assumes that 3D object surface can be constructed by some transformations, e.g., translation, rotation, of a plane curve. 
    Then, it adopts the notion of RHM to derive a general tracking framework. 
    This approach necessitates a special formulation of the recursive estimator in accordance with the particular transformation considered. 
    However, the required prior information about the object shape may not be available for a standard tracking application. 
    
    Another branch of studies avoids imposing a parametric shape description; instead it is based on a point cloud representation of the shape, which is formed by collecting measurements over time, \cite{moosmann2013joint, held2016robust, kraemer2017simultaneous}. 
    These methods facilitate joint tracking and shape learning of arbitrary objects in the presence of continuously available, high-precision and informative measurements. 
    However, as they do not feature a principled representation of the underlying shape, there arise robustness issues with the sparsity of the measurements, for example, due to increasing distance, change of the vantage point and occlusions. 
    In addition, the storage and the computational requirements scale with the size of the object extent.
    
    In this study, we consider the problem of tracking 3D objects while simultaneously learning their shapes using 3D point cloud data. 
    3D measurements carry substantial information such that they not only convey clues about the kinematics but also reveal characteristics of the object extent naturally. 
    However, estimating unknown shapes from noisy point cloud data is a challenging task, and the problem gets even more severe when the objects are in motion. 
    This is mainly due to the inherent interdependence between the pose and the shape description. 
    Therefore, the reliable estimation of one necessitates precise information about the other. 
    With this in mind, we formulate the problem as the joint estimation of both kinematic and shape variables in a unified framework. 
    For the description of the object extent, we adopt a GP-based approach that facilitates a flexible representation with favorable analytical properties. 
    In this paper, we significantly extend our previous work on 3D object tracking, \cite{kumru20183d}. 
    In particular, the contributions can be listed as follows. 
    \begin{itemize}
    	\item We propose two novel probabilistic representations for 3D extent. 
    	The first one expresses the unknown 3D surface by a radial function in spherical coordinates. 
    	The second exploits the correspondence between a 3D shape and its projections onto multiple planes, and thus characterizes the original 3D surface by a collection of 2D contours of these projections. 	
    	The probabilistic shape representation is achieved by modeling the above descriptions by GPs without imposing any parametric form. 
    	By doing so, we attain a flexible basis to estimate the shape of a wide range of objects.
    	
    	\item This approach enables us to treat the unknown extent within the Bayesian framework such that the posterior distribution offers an analytical expression for the object shape with well-defined confidence intervals, and any available prior information can be incorporated easily. 
    	
    	\item We develop measurement models to express the relation between the point measurements and the object extent using an efficient approximation of the GP regression. 
    	
    	\item The kinematics and the object extent are efficiently inferred by an extended Kalman filter regarding a unified state space model. 
    	
    	\item The orientation of the object is described by the unit quaternions, and a novel rotational motion model is derived for effective estimation of the full 3D orientation and the angular rates. 
    	
    	\item The performance of the suggested algorithms is comprehensively evaluated using both simulated and real measurements. 	
    \end{itemize}
	
	\section{Extent Model for 3D Objects} \label{sec: ExtentModel}
    Different types of sensors provide different level of details about objects of interest. 
    One particular group of sensors, such as depth cameras and LIDARs, generates 3D point cloud data that capture salient characteristics of their environment. 
    It is an open research question as to how we can harness the full potential of the available information. 
    Other than its potential benefits in tracking such as clutter rejection and data association; a detailed shape estimate can provide valuable insight for the classification of the object, which in turn may prove useful to anticipate future behavior. 
    With this improved perception capabilities, it might be possible to develop tailored ways of interaction with the environment.
    
    To accomplish effective shape learning by processing 3D point cloud data, a suitable description of the object extent needs to be formulated that meets the following specifications. 
    First, it is required to have high representational power to be able to apply to a wide range of objects with various 3D shapes. 
    In addition, it should be sufficiently compact so that it will enable an efficient online tracking algorithm. 
    
\begin{figure}[t]
    	\centering
    	\includegraphics{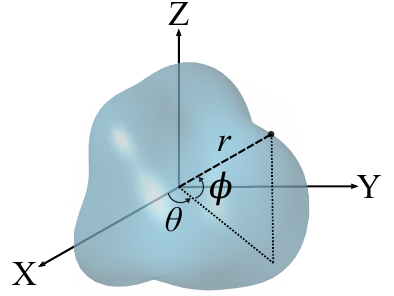}
    	\caption{Object extent description in spherical coordinates. }
    	\centering
    	\label{fig:kumru1}
    \end{figure}

    In this regard, we model the object shape in spherical coordinates by means of a radial function $f(\theta, \phi)$. 
    The arguments of this function are the azimuth, ${\theta \in [-\pi, \pi]}$, and the elevation angles, ${\phi \in [-\frac{\pi}{2}, \frac{\pi}{2}]}$, and the output, $r$, is the distance between the center of the object and the point on the surface at the corresponding spherical angle pair, i.e., ${r = f(\theta, \phi)}$. 
    Fig. \ref{fig:kumru1} illustrates the representation for an example object. 
    
    This representation summarizes the 3D shape exclusively by the external boundary (surface) of the object, considering that the point cloud measurements are merely originated from the surface. 
    Additionally, it implicitly assumes that the latent shape is star-convex\footnote{A set $\mathcal{S}$ is star-convex with respect to the origin if each line segment from the origin to any point in $\mathcal{S}$ is fully contained in $\mathcal{S}$.}. 
    This assumption does not introduce a strict limitation as star-convex shapes present an adequately broad class for object tracking applications. 
    
    The main aim of the upcoming sections is to construct a unified state space model to serve as a basis for the joint estimation of the kinematics and the extent of the object. 
    The corresponding state vector includes both the kinematics and a parametric description of the given extent model. 
    This description will be obtained by developing a proper GP model for the radial function. 
    More specifically, we precisely construct the GP model such that it effectively accounts for the inherent spatial correlation within the object surface. 
    Furthermore, we utilize a recursive approximation of GP modeling to avoid associated computational difficulties. 
    The resulting algorithm accomplishes a probabilistic representation of the latent extent in a principled manner.
    Besides, it is able to maintain the local uncertainty information of the extent which becomes vital for robust tracking and shape learning in scenarios including occlusions and sparse sampling. 
    
    The next section first briefly introduces the standard GP regression and then elaborates on its recursive approximation.
	
    \section{Gaussian Processes} \label{sec:GP}
    A Gaussian Process (GP) is a stochastic model which specifies a probability distribution in the function space for a function $f(\cdot)$, \cite{rasmussen2006gaussian}. 
    We hereby engage a GP to model the radial function expressing the unknown extent of the object. 
    A GP is uniquely defined by the mean $\mu(u)$  and the covariance function $k(u,u')$ defined as\footnote{Considering the radial distance function, the definition of the GP model is deliberately given for a scalar-valued function.}
    \begin{subequations}
    	\begin{align}
    	\mu(u) &= \mathbb{E}[f(u)], \\
    	k(u,u') &= \mathbb{E}[(f(u)-\mu(u))(f(u')-\mu(u'))].
    	\end{align}
    \end{subequations}
    The corresponding GP model is denoted as
    \begin{equation}
    f(u) \sim \mathcal{GP}(\mu(u), k(u,u')), \nonumber
    \end{equation}
    where $u$ is the input of the function, which is specified as a scalar for notational simplicity. 
    This model will directly generalize to vector inputs as discussed in Section \ref{sec:GPModelofExtent}.
    
    A GP can also be interpreted as a collection of random variables, any finite number of which have a joint Gaussian distribution that is consistent with the specified mean and covariance functions. 
    The joint distribution of the function evaluations at the inputs, $u_1,...,u_N$,  reads as
    \begin{subequations} \label{eq:descriptionGP}
    	\begin{align}
    	\begin{bmatrix}
    	f(u_1)\\
    	\vdots\\
    	f(u_N)
    	\end{bmatrix} 
    	\sim
    	\mathcal{N}(\boldsymbol{\mu}, K), \ \text{where}\ 
    	\boldsymbol{\mu} = \begin{bmatrix}
    	\mu(u_1)\\
    	\vdots\\
    	\mu(u_N)
    	\end{bmatrix},
    	\end{align}	
    	and
    	\begin{equation}
    	K = 
    	\begin{bmatrix}
    	k(u_1, u_1) &\dots & k(u_1, u_N)\\
    	\vdots &   & \vdots\\
    	k(u_N, u_1) &\dots & k(u_N, u_N)
    	\end{bmatrix}.
    	\end{equation}
    \end{subequations}
    
    \subsection{Gaussian Process Regression}
    Prior belief about the unknown function encoded by the GP can be conveniently conditioned on the information provided by observations. 
    For this purpose, a noisy observation $m$ can be described as the true function output perturbed by an independent Gaussian noise $e$, 
    \begin{equation} \label{eq:measModelGP}
    m = f(u) + e, \ \ e \sim \mathcal{N}(0,R). 
    \end{equation}
    
    Assume that we seek for the refined distribution of the function values ${\mathbf{f} \triangleq [{f(u^\mathbf{f}_1})\ \dots\ f(u^\mathbf{f}_{N^\mathbf{f}})]^\top}$ at the inputs ${\mathbf{u}^\mathbf{f} \triangleq [u^\mathbf{f}_1\ \dots\ u^\mathbf{f}_{N^\mathbf{f}}]^\top}$. 
    Available measurements are denoted by ${\mathbf{m} \triangleq [m_1\ \dots\ m_N]^\top}$ which are originated from the inputs ${\mathbf{u} \triangleq [u_1\ \dots\ u_N]^\top}$. 
    The GP model together with the measurement model \eqref{eq:measModelGP} leads to the following joint distribution, 
    \begin{subequations}
    	\begin{align}
    	\begin{bmatrix} 
    	\mathbf{m}\\ 
    	\mathbf{f}
    	\end{bmatrix} \sim \mathcal{N}
    	\left(
    	\begin{bmatrix} 
    	\boldsymbol{\mu}(\mathbf{u})\\ 
    	\boldsymbol{\mu}(\mathbf{u}^\mathbf{f})
    	\end{bmatrix},
    	\begin{bmatrix}
    	K(\mathbf{u}, \mathbf{u})+ I_N R &  K(\mathbf{u}, \mathbf{u}^\mathbf{f})\\
    	K(\mathbf{u}^\mathbf{f}, \mathbf{u}) & K(\mathbf{u}^\mathbf{f}, \mathbf{u}^\mathbf{f})
    	\end{bmatrix}
    	\right),
    	\end{align}	
    	\text{where}\
    	\begin{align}
    	\begin{split}
    	    \boldsymbol{\mu}(\mathbf{u}) &= [\mu(u_1)\ \dots\ \mu(u_N)]^\top, \\
    	    \boldsymbol{\mu}(\mathbf{u}^\mathbf{f}) &= [\mu(u^\mathbf{f}_1)\ \dots\ \mu(u^\mathbf{f}_{N^\mathbf{f}})]^\top, \\
        	K(\mathbf{u}, \mathbf{u}^\mathbf{f}) &= 
        	\begin{bmatrix}
        	k(u_1, {u^\mathbf{f}_1}) &\dots & k(u_1, u^\mathbf{f}_{N^\mathbf{f}})\\
        	\vdots & \  & \vdots\\
        	k(u_N, u^\mathbf{f}_1) &\dots & k(u_N, u^\mathbf{f}_{N^\mathbf{f}})
        	\end{bmatrix},
    	\end{split}
    	\end{align}
    \end{subequations}
    $I_N$ indicates an \textit{N}-by-\textit{N} identity matrix. 
    
    For the sake of brevity, here we will present the case where the mean function of the GP model is zero. However, the method can easily be generalized to an arbitrary mean function and the relevant modifications are presented in Appendix A. Under this assumption, the conditional distribution $p(\mathbf{f}|\mathbf{m})$ can be expressed as 
    \begin{subequations} \label{eq:GPRegression}
    	\begin{equation}
    	p(\mathbf{f}|\mathbf{m}) \sim \mathcal{N}(A\mathbf{m}, P),
    	\end{equation}
    	where
    	\begin{align}
    	A &= K(\mathbf{u}^\mathbf{f}, \mathbf{u}) {K}^{-1}_y, \\
    	P &= K(\mathbf{u}^\mathbf{f}, \mathbf{u}^\mathbf{f}) - K(\mathbf{u}^\mathbf{f}, \mathbf{u}) {K}^{-1}_y K(\mathbf{u}, \mathbf{u}^\mathbf{f}),\\
    	K_y &= K(\mathbf{u}, \mathbf{u}) + I_N  R. \label{eq:GPEstimate2}
    	\end{align}
    \end{subequations}
    
    \subsection{Recursive Gaussian Process Regression}
    The GP regression necessitates to process all available information in a single batch as the complete measurement vector $\mathbf{m}$ and the corresponding covariance matrix ${K}_y$ appear in \eqref{eq:GPRegression}. 
    While this attribute can be interpreted to be the primary strength of GP modeling since it enables to draw conclusions directly from the observations, it also poses some computational problems for certain settings. 
    Specifically, in object tracking, the aim is to compute the posterior density $p(\mathbf{f}|m_{1:k})$ at time $k$ using measurements which are acquired sequentially in time. 
    For this problem, online inference can be achieved by a recursive algorithm which efficiently updates the posterior by considering the newly available measurements. 
    In this respect, the standard GP regression is not applicable due to its increasing needs for computational sources and memory storage with the accumulation of measurements over time. 
    Therefore, we hereby rely on an approximation of the GP, that basically summarizes the original model at a finite set of basis points. 
    The approximation was initially proposed in \cite{huber2013recursive,huber2014recursive}, and then applied to the object tracking problem in \cite{wahlstrom2015extended}. 
    
    In this approach, the objective is to derive a formulation of the posterior distribution $p(\mathbf{f}|m_{1:N})$ that enables recursive implementation. 
    To this end, the posterior is first expanded as the collection of the following terms by applying the Bayes' law iteratively. 
    \begin{subequations}
    	\begin{align}
    	p(\mathbf{f}|m_{1:N}) &\propto p(m_N|\mathbf{f}, m_{1:N-1}) p(\mathbf{f}|m_{1:N-1}) \\
    	&\propto  p(\mathbf{f}) p(m_1|\mathbf{f}) \prod_{k=2}^{N} p(m_k|\mathbf{f},m_{1:k-1})
    	\end{align}
    \end{subequations}
    
    \textbf{\textit{Assumption:}} $\mathbf{f}$ provides the sufficient statistics for $m_k$. 
    
    \noindent Under this assumption, $m_k$ conditioned on $\mathbf{f}$ becomes independent from all previous measurements, $m_{1:k-1}$, i.e., 
    \begin{align}
    p(m_k|\mathbf{f}, m_{1:k-1}) \approx p(m_k|\mathbf{f}).
    \end{align}
    Notice that the assumption becomes exact if the inputs of $m_k$ form a subset of the inputs of $\mathbf{f}$. 
    Moreover, it can be claimed to be a reasonable approximation when the distance between the inputs of $m_k$ and the inputs of $\mathbf{f}$ is sufficiently small compared to the characteristic length-scale of the covariance function. 
    In this study, we want to model the unknown radial function whose input is the spherical angle pair. 
    As the set of the possible input values has a well-defined boundary, it is possible to sufficiently sample this set by a finite number of basis points which will be located equidistantly. 
    
    The above assumption leads to a setting where we essentially treat $\mathbf{f}$ to be the latent variable and the measurements provide noisy observations of it. 
    Accordingly, once the measurement likelihood and the initial prior densities are defined, it is possible to apply recursive Bayesian inference for $\mathbf{f}$. 
    With this in mind, we simply refer to the underlying GP model to offer these densities in a principled way. 
    At first, the joint distribution of the measurement $m_k$ and $\mathbf{f}$ is revealed by the definition of the GP as  
    \begin{equation}
    \begin{bmatrix} 
    m_k\\ 
    \mathbf{f}
    \end{bmatrix} \sim \mathcal{N}
    \left(
    \mathbf{0},
    \begin{bmatrix}
    k(u_k, u_k)+R &  K(u_k, \mathbf{u}^\mathbf{f})\\
    K(\mathbf{u}^\mathbf{f}, u_k) & K(\mathbf{u}^\mathbf{f}, \mathbf{u}^\mathbf{f})
    \end{bmatrix}
    \right).
    \end{equation}
    Then, the joint distribution immediately offers the following likelihood and prior densities, 
    \begin{subequations}  \label{eq:recursiveGPFormula}
    	\begin{align}
    	p(m_k|\mathbf{f}) &= \mathcal{N}(m_k; H_k^\mathbf{f}\mathbf{f}, R_k^\mathbf{f}), \\
    	p(\mathbf{f}) &= \mathcal{N}(\mathbf{0}, P_0^\mathbf{f}),
    	\end{align}
    	where
    	\begin{align}
    	H^\mathbf{f}_k &= H^\mathbf{f}(u_k) =  K(u_k, \mathbf{u}^\mathbf{f}) [K(\mathbf{u}^\mathbf{f}, \mathbf{u}^\mathbf{f})]^{-1}, \\
    	R^\mathbf{f}_k &= R^\mathbf{f}(u_k) = k(u_k, u_k)+R \nonumber\\
    	& \quad - K(u_k, \mathbf{u}^\mathbf{f}) [K(\mathbf{u}^\mathbf{f}, \mathbf{u}^\mathbf{f})]^{-1} K(\mathbf{u}^\mathbf{f}, u_k),\\
    	P_0^\mathbf{f} &= K(\mathbf{u}^\mathbf{f}, \mathbf{u}^\mathbf{f}).
    	\end{align}
    \end{subequations}
    
    The structure of \eqref{eq:recursiveGPFormula} allows us to construct the following state space model to which a standard Kalman filter can be applied for recursive inference, \cite{wahlstrom2015extended}. 
    \begin{subequations} \label{eq:extentStateSpace}
    	\begin{align}
    	\mathbf{f}_{k+1} &= \mathbf{f}_{k}, \label{eq:dyn_mdl_extent}\\
    	m_k &= H^\mathbf{f}_k\ \mathbf{f}_{k} + e_k^\mathbf{f},\quad e_k^\mathbf{f} \sim \mathcal{N}(0, R^\mathbf{f}_k), \\
    	\mathbf{f}_0 &\sim \mathcal{N}(\mathbf{0}, P_0^\mathbf{f}), 
    	\end{align}
    \end{subequations}
    where $\mathbf{f}_{k}$ is basically defined as the latent function values evaluated at the predetermined inputs, i.e., $\mathbf{f}_{k} \triangleq \mathbf{f}$.
    
    The benefits of having such a state space model for the object extent are twofold: 
    first, we can easily engage a dynamical model in \eqref{eq:dyn_mdl_extent} to express the time evolution of the extent; 
    second, it can simply be augmented by another state space model to obtain a unified representation. 
    We will basically benefit from these advantages while developing the unified state space model in Section \ref{sec:GPETT3DStateSpaceModel}.
	
	\section{GP Modeling of Object Extent} \label{sec:GPModelofExtent}
    In this section, the radial function which expresses the object extent is to be modeled via a GP. 
    By doing so, we will be able to facilitate effective shape learning in the probabilistic framework by using incomplete and noisy point measurements. 
    
    GP lends itself conveniently to extent modeling since it is naturally able to describe the spatial correlation between different sections the object surface. 
    In addition, it maintains local uncertainty information associated with the object surface which is critical for accurate gating and association of the measurements leading to robust tracking performance. 
    
    A GP model is uniquely defined by its mean and covariance functions, hence the main focus of this discussion is to rigorously construct these functions regarding the characteristics of the extent representation. 
    Note that we hereby put forward a generic approach to be able to apply to arbitrarily shaped objects; however, prior knowledge about the object shape can also be systematically incorporated by adjusting these functions accordingly, e.g., see \cite{martens2017geometric}. 
    
    As discussed earlier, the output of the radial function is the distance $r$, and the input is the pair of azimuth and elevation angles $(\theta, \phi)$, i.e., $r = f(\theta, \phi)$. 
    For notational simplicity, the pair $(\theta, \phi)$ is assigned to $\boldsymbol{\gamma}$, i.e., ${\boldsymbol{\gamma} \triangleq (\theta, \phi)}$ and ${r = f(\boldsymbol{\gamma})}$. 
    Therefore, we denote the mean and covariance functions as $\mu (\boldsymbol{\gamma})$ and $k(\boldsymbol{\gamma}, \boldsymbol{\gamma}')$, respectively, and ${f(\boldsymbol{\gamma}) \sim \mathcal{GP}(\mu (\boldsymbol{\gamma}), k(\boldsymbol{\gamma}, \boldsymbol{\gamma}'))}$ indicates the GP model. 
    
    \subsection{Mean Function}
    The mean function of the GP model is assumed to be an unknown constant having a normal distribution, i.e., 
    \begin{equation} \label{eq:meanDer1}
    	\mu (\boldsymbol{\gamma}) = r, \quad \text{where}\ r \sim \mathcal{N}(\mu_r, \sigma^2_r).
    \end{equation}
    Additionally, we can obviously express the original GP model as in
    \begin{align} \label{eq:meanDer2}
    f(\boldsymbol{\gamma}) = \bar{f}(\boldsymbol{\gamma}) + \mu (\boldsymbol{\gamma}),\ \text{where}\ \bar{f}(\boldsymbol{\gamma}) \sim \mathcal{GP}(0, k(\boldsymbol{\gamma}, \boldsymbol{\gamma}')).
    \end{align} 
    Consequently, by using the prior distribution of $\mu (\boldsymbol{\gamma})$ in \eqref{eq:meanDer1}, we can obtain an equivalent representation of \eqref{eq:meanDer2} as follows, \cite[Ch. 2.7]{rasmussen2006gaussian}, \cite{o1978curve}. 
    \begin{align} \label{eq:totCov}
    	f(\boldsymbol{\gamma}) \sim \mathcal{GP}(\mu_r, &k_{total}(\boldsymbol{\gamma}, \boldsymbol{\gamma}')), \nonumber\\
    	\text{where}\quad &k_{total}(\boldsymbol{\gamma}, \boldsymbol{\gamma}') = k(\boldsymbol{\gamma}, \boldsymbol{\gamma}') + \sigma^2_r. 
    \end{align}
    
    \subsection{Covariance Function}
    Selection of the covariance function for a GP is of great importance since it basically determines the characteristics of the functions to be learned. 
    In this application, it is required to conform to the fundamentals of 3D object geometry as it encodes the spatial correlation between points on the extent. 
    
    The design of the covariance function is initiated from the exponentiated quadratic function, which is accepted to be the de facto choice in various fields, \cite{rasmussen2006gaussian}, 
    \begin{equation} \label{eq:SEcov1}
    	k(\boldsymbol{\gamma}, \boldsymbol{\gamma}') = \sigma_f^2 e^{-\frac{d^2(\boldsymbol{\gamma}, \boldsymbol{\gamma}')}{2l^2}}, 
    \end{equation}
    where $\sigma_f^2$ stands for the prior variance, $l$ is the length-scale and $d(\boldsymbol{\gamma}, \boldsymbol{\gamma}')$ calculates the relative distance between two inputs. 
    The prior variance specifies the typical amount of variation observed among the functions sampled from a GP. 
    Accordingly, increasing prior variance in the proposed extent model would imply that there is relatively less prior information about the size of the objects to be tracked.  
    Additionally, the length scale determines the `smoothness' of the functions modeled by a GP. 
    In this context, a shorter length-scale would suggest that the corresponding extent is potentially sharper (exhibiting high-frequency oscillations) while increasing the length-scale would lead to a smoother extent representation. 
    
    The unconventional aspect of the employed covariance function is the formulation of the distance, $d(\boldsymbol{\gamma}, \boldsymbol{\gamma}')$. 
    It is specified to imply higher correlation for closer regions compared to those which are rather separated. 
    An immediate option for the distance definition could be the Euclidean distance, i.e., ${d(\boldsymbol{\gamma}, \boldsymbol{\gamma}') = \|\boldsymbol{\gamma} - \boldsymbol{\gamma}' \|}$, as used in \cite{wahlstrom2015extended}. 
    However, being inconsistent with the basics of the spherical geometry, it leads to erroneous correlation patterns for the extent defined in the spherical coordinates. 
    As a simple example, consider $\boldsymbol{\gamma} = \left(0, \frac{\pi}{2}\right)$ and $\boldsymbol{\gamma}' = \left(\pi, \frac{\pi}{2}\right)$ both pointing to the upper pole of a sphere. 
    For these inputs, the Euclidean distance is computed as $\pi$ which is also equal to the distance for any two spherical angles pointing to opposite directions, e.g., the upper and the lower poles, i.e., $\boldsymbol{\gamma} = \left(0, \frac{\pi}{2}\right)$ and $\boldsymbol{\gamma}' = \left(0, -\frac{\pi}{2}\right)$. 
    Such problems encountered in the angular estimation applications are commonly addressed by suitable angular distance measures in the literature (see, for example, \cite{kurz2016recursive} and the references therein). 
    In this study, we suggest an alternative distance definition that naturally induces a proper correlation structure. 
    Consider two unit-length vectors expressed by the spherical angles pairs, $\boldsymbol{\gamma}$ and $\boldsymbol{\gamma}'$. 
    We set $d(\boldsymbol{\gamma}, \boldsymbol{\gamma}')$ to be the angle between these two vectors. 
    From another viewpoint, this definition is the angle corresponding to the shortest arc on a sphere that connects the two points described by $\boldsymbol{\gamma}$ and $\boldsymbol{\gamma}'$. 
    The analytical expression for the definition can be written as 
    \begin{multline} \label{eq:SEcov1_2}
    	d(\boldsymbol{\gamma}, \boldsymbol{\gamma}') = \arccos \big(\cos(\phi)\cos(\phi ')\cos(\theta)\cos(\theta ') \\ + \cos(\phi)\cos(\phi ')\sin(\theta)\sin(\theta ') + \sin(\phi)\sin(\phi ') \big), 
    \end{multline}
    where $\boldsymbol{\gamma} = (\theta, \phi)$ and $\boldsymbol{\gamma}' = (\theta ', \phi ')$. 
    Notice that with this formulation, the distance takes values within the interval $[0, \pi]$, and any coincident angle pair is mapped to $0$ while opposite directions compute $\pi$. 
    
    Finally, the total covariance function in \eqref{eq:totCov} is attained as 
    \begin{align} \label{eq:SEcov2}
    	k_{total}(\boldsymbol{\gamma}, \boldsymbol{\gamma}') &= k(\boldsymbol{\gamma}, \boldsymbol{\gamma}') + \sigma_r^2, \nonumber \\
    	&= \sigma_f^2 e^{-\frac{d^2(\boldsymbol{\gamma}, \boldsymbol{\gamma}')}{2l^2}} + \sigma_r^2.
    \end{align}
    
    The formulation of the covariance function can easily be adapted regarding the characteristics of a specific application. 
    For instance, let us consider a setting where an autonomous robotic manipulator is assigned to grasp the surrounding objects with unknown shapes. 
    If there is prior information available regarding the environment, e.g., the objects of interest are symmetric around their azimuth axis, then one can account for this specification by adjusting the utilized distance function accordingly. 
	
	\section{State Space Model} \label{sec:GPETT3DStateSpaceModel}
    In this section, we will develop a state space model to be used in object tracking. 
    This model is based on the state vector involving both the kinematics and the extent representation of the object. 
    In this setting, joint estimation of this aggregated state variables will be accomplished by a single inference algorithm. 
    In other words, the idea of leveraging the latent shape information for object tracking is basically realized by this formulation. 
    
    The state vector is defined as ${\mathbf{x}_{k} \triangleq \left[\bar{\mathbf{x}}_{k}^\top \ \mathbf{f}_k^\top \right]^\top}$ where $\bar{\mathbf{x}}_{k}$ consists of the translational, ${\mathbf{x}_{k}^t}$, and the rotational, ${\mathbf{x}_{k}^r}$, kinematic variables, i.e., ${\bar{\mathbf{x}}_{k}^\top \triangleq \left[ {\mathbf{x}_{k}^t}^\top \ {\mathbf{x}_{k}^r}^\top \right]^\top}$, and $\mathbf{f}_k$ indicates the extent representation. 
    
\begin{figure}[b]
    	\centering
    	\includegraphics{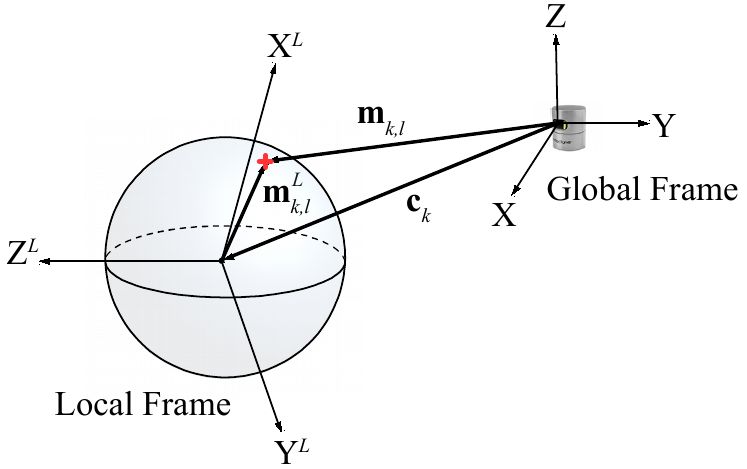}
    	\caption{Illustration of the coordinate frames and the vectors regarded in the measurement model. }
    	\centering
    	\label{fig:kumru2}
    \end{figure}

    The definition of the state vector makes use of two distinct coordinate frames as shown in Fig. \ref{fig:kumru2}. 
    The first one is the global coordinate frame which is fixed to the sensor; the second one is the local coordinate frame which is attached to the object to be tracked. 
    As the local frame performs exactly the same motion with the object, it allows to describe the extent in a consistent manner. 
    Accordingly, the extent information is maintained in the local coordinate frame while the object motion is estimated in the global coordinate frame. 
    
    An overview of the state space model is given by the following set of equations, 
    \begin{subequations} \label{eq:StateSpaceModel}
    	\begin{align}
    	\mathbf{x}_{k+1} &= F_k \mathbf{x}_{k} + \mathbf{w}_k,\quad \mathbf{w}_k \sim \mathcal{N}(\mathbf{0}, Q_k), \label{eq:StateSpaceModel1}\\
    	\mathbf{0} &= \mathbf{h}(\mathbf{x}_k, \mathbf{m}_{k,l}) + \mathbf{e}_{k,l}, \quad \mathbf{e}_{k,l} \sim \mathcal{N}(\mathbf{0}, R_{k,l}), \label{eq:StateSpaceModel_meas} \\
    	\mathbf{x}_0 &\sim \mathcal{N}(\boldsymbol{\mu}_0, P_0),  \label{eq:StateSpaceModel3}
    	\end{align}
    \end{subequations}
    where $k$ is the time index;  $\mathbf{w}_k$ and $\mathbf{e}_{k,l}$ indicate the zero-mean process and the measurement noise, respectively. 
    These are assumed to be Gaussian with covariance matrices defined as ${\text{cov}[\mathbf{w}_k] = Q_k}$ and ${\text{cov}[\mathbf{e}_{k,l} ] = R_{k,l}}$. 
    The following subsections will introduce the details of these equations starting from the derivation of the process model.

    \subsection{Process Model} \label{sec:GPETT3DProcessModel}
    The overview of the process model, which describes the evolution of the states over time, is given by the linear Gaussian model in \eqref{eq:StateSpaceModel1}. 
    Recall that the state vector is formed by concatenating the kinematics and the extent representation, i.e., ${\mathbf{x}_{k} \triangleq \left[\bar{\mathbf{x}}_{k}^\top \ \mathbf{f}_k^\top \right]^\top}$. 
    It is assumed that the dynamics of these two state components do not interact with each other. 
    Therefore, the process model includes two independent subsystems and can explicitly be written as 
    \begin{align}
    F_k &= \begin{bmatrix}
    \bar{F}_k & 0 \\
    0 & F^\mathbf{f}
    \end{bmatrix}, \ 
    Q_k = \begin{bmatrix}
    \bar{Q}_k & 0 \\
    0 & Q_k^\mathbf{f}
    \end{bmatrix}. 	
    \end{align}
    Accordingly, the prior distribution of the state given in \eqref{eq:StateSpaceModel3} can be specified by
    \begin{align}
    \boldsymbol{\mu}_0 &= \begin{bmatrix}
    \bar{\boldsymbol{\mu}}_0 \\
    \boldsymbol{\mu}^\mathbf{f}_0
    \end{bmatrix}, \ 
    \ P_0 = \begin{bmatrix}
    \bar{P}_0 & 0 \\
    0 & P^\mathbf{f}_0
    \end{bmatrix}.
    \end{align}
    
    Our formulation does not put any restriction on the selection of the process model; hence one can freely design the model for both the kinematics and the extent accounting for the characteristics of a specific application.
    
    We hereby use the following dynamical model for the object extent:
    \begin{subequations} \label{eq:extentDyn_maxent}
    	\begin{align}
    	\mathbf{f}_{k+1} &= \mathbf{f}_{k} + \mathbf{w}_k,\quad \mathbf{w}_k \sim \mathcal{N}(\mathbf{0}, Q_k^\mathbf{f}),
    	\end{align}
    	where
    	\begin{align}
    	Q_k^\mathbf{f} &= \left( \frac{1}{\lambda} - 1\right) P_{k|k}^\mathbf{f}, 
    	\end{align}
    \end{subequations}
    which implies that $F^\mathbf{f}$ is set to be an identity matrix. 
    $P_{k|k}^\mathbf{f}$ denotes the covariance of the estimated extent state. 
    Notice that the prediction density computed using this model has the same mean with that of the estimated density, while the  prediction covariance is scaled up as $P_{k+1|k}^\mathbf{f} = \frac{1}{\lambda} P_{k|k}^\mathbf{f}$ for $\lambda<1$.   
    The model provides a maximum entropy distribution for the extent prediction, as shown in \cite{ozkan2013marginalized}. 
    This approach reflects that the transition density is unknown, while the Kullback-Leibler divergence to the prediction density is upper bounded. 

    With this model, we can basically account for the possible changes in the object extent. 
    Therefore, it potentially facilitates the tracking of nonrigid objects. 
    In addition, it enables recovery from erroneously integrated shape information, which might occur due to temporal errors in the pose estimates, especially during the initial phases of the tracking. 
    
    In this study, we assume that the rotational and the translational motion can be modeled independently. 
    Therefore,  $\bar{F}_k$ and $\bar{Q}_k$ can be written as 
    \begin{align} \label{eq:DTStateMatrices}
    \bar{F}_k &= \begin{bmatrix}
    F^t & 0 \\
    0 & F_k^r
    \end{bmatrix}, \ 
    \bar{Q}_k = \begin{bmatrix}
    Q^t & 0 \\
    0 & Q_k^r
    \end{bmatrix}. 	
    \end{align}
    
    \subsubsection{Translational Motion Model}
    We employ the well-known almost constant velocity model for the translational motion as given in \eqref{eq:TransCCModel}. 
    The translational kinematic state is defined as ${\mathbf{x}_{k}^t \triangleq \left[ \mathbf{c}_k^\top \ \mathbf{v}_k^\top \right]^\top}$ where ${\mathbf{c}_k \in \mathbb{R}^3}$ is position of the object center, and ${\mathbf{v}_k \in \mathbb{R}^3}$ stands for the velocity of the center. 
    \begin{subequations} \label{eq:TransCCModel}
    	\begin{align}
    	\mathbf{x}_{k+1}^t &= F^t \mathbf{x}_{k}^t + \mathbf{w}_k^t,\quad \mathbf{w}_k^t \sim \mathcal{N}(\mathbf{0}, Q^t), \\
    	\mathbf{x}_0^t &\sim \mathcal{N}(\boldsymbol{\mu}_0^t, P_0^t),
    	\end{align}
    	where 
    	\begin{align}
    	F^t &= \begin{bmatrix}
    	1 & T \\
    	0 & 1
    	\end{bmatrix} \otimes I_3,\ 
    	Q^t = \begin{bmatrix}
    	\frac{T}{3}^3 & \frac{T}{2}^2 \\[1pt]
    	\frac{T}{2}^2 & T
    	\end{bmatrix} \otimes (\sigma^2_c I_3).
    	\end{align}
    \end{subequations}
    $\sigma^2_c$ is the process noise variance for the center, $\otimes$ is the Kronecker product, and $T$ is sampling time. 
    
    \subsubsection{Rotational Motion Model} \label{sec:RotationalMM}
    The representation of the orientation of the object is of paramount importance as it directly affects the performance of tracking and shape estimation.
    In the literature, there are many alternative representations for the orientation of 3D objects. 
    The main challenges associated with these are the included singularities and the inherent constraints, \cite{crassidis2007survey}. 
    For example, expressing the orientation using three variables, Euler angles offer a minimal representation. 
    However, this approach is impeded by the singularities, and it can not provide a global orientation description. 
    On the other hand, the four-component unit quaternion presents the lowest dimensional representation that avoids any possible singularity. 
    Nevertheless, it introduces a nonlinear equality constraint such that the norm of the quaternion has to be always exactly one. 
    Hence, it requires special treatment during the estimation process. 
    
    Considering the aforementioned difficulties, we rely on an alternative parametrization of the orientation. 
    This representation stores a reference orientation as a unit quaternion, and the orientation deviation from this reference is expressed by a three-component error vector. 
    By this approach, a standard extended Kalman filter (EKF) can easily be employed to estimate the error vector, and the reference orientation is periodically updated by the estimated deviation. 
    The resulting algorithm is referred to as multiplicative EKF (MEKF), \cite{markley2003attitude}. 
    The virtue of the method is twofold: first, it can represent the orientation globally as it essentially exploits a unit quaternion description; second, the norm constraint of the unit quaternion is naturally satisfied as the quaternion gets updated appropriately by the error vector.
    
    In the rest of this subsection, we will rely on this representation to derive a constant velocity model for the object orientation. 
    The notation and the basics of the quaternion algebra is adopted from \cite{maeder2011attitude}, which closely follows \cite{markley2004attitude}.
    
    The unit quaternion ${\mathbf{q} \in \mathbb{R}^4}$ is defined as 
    \begin{align}
    \mathbf{q} \triangleq \begin{bmatrix}
    q_1 \ q_2 \ q_3 \ q_4
    \end{bmatrix}^\top = 
    \begin{bmatrix}
    \bar{\mathbf{q}}^\top \ q_4 
    \end{bmatrix}^\top,
    \end{align}
    where $\bar{\mathbf{q}} \triangleq [q_1\ q_2\ q_3]^\top$ and 
    ${|\mathbf{q}|^2 = |\bar{\mathbf{q}}|^2 + q_4^2 = 1}$. 
    
    Then, the rotation matrix $R_G^L(\mathbf{q})$ that expresses the orientation of the local frame with respect to the global frame is given by
    \begin{equation}
    R_L^G(\mathbf{q}) = (q_4^2 - \bar{\mathbf{q}}^\top \bar{\mathbf{q}}) I_3 + 2 \bar{\mathbf{q}} \bar{\mathbf{q}}^\top - 2 q_4 [\bar{\mathbf{q}} \times]\ ,
    \end{equation}
    where $[\bar{\mathbf{q}} \times]$ is the cross product matrix 
    \begin{equation} 
    [\bar{\mathbf{q}} \times] = 
    \begin{bmatrix}
    0 & -q_3 & q_2 \\
    q_3 & 0 & -q_1 \\
    -q_2 & q_1 & 0
    \end{bmatrix}.
    \end{equation}
    The quaternion product is defined as 
    \begin{align} \label{eq:quatProd}
    \mathbf{p} \odot \mathbf{q} =
    \begin{bmatrix}
    p_4 \bar{\mathbf{q}} + q_4 \bar{\mathbf{p}} - \bar{\mathbf{p}} \times \bar{\mathbf{q}} \\
    p_4 q_4 - \bar{\mathbf{p}}^\top \bar{\mathbf{q}}
    \end{bmatrix},
    \end{align}
    which leads to the following useful property
    \begin{align}
    R(\mathbf{p}) R(\mathbf{q}) = R (\mathbf{p} \odot \mathbf{q})\ ,
    \end{align}
    where $R(\mathbf{p})$ and $R(\mathbf{q})$ denote the rotation matrices corresponding to the unit quaternions $\mathbf{p}$ and $\mathbf{q}$, respectively. 
    This equation reveals that successive rotation operations can be expressed in terms of the quaternion product. 
    Based on this fact, we can describe the orientation of the object as follows
    \begin{align} \label{eq:quatRef}
    \mathbf{q} = \delta\mathbf{q}(\mathbf{a}) \odot \mathbf{q}_\text{ref}\ ,
    \end{align}
    where $\mathbf{q}_\text{ref}$ indicates a reference orientation, and $\delta\mathbf{q}(\cdot)$ corresponds to the deviation from the reference. 
    $\delta\mathbf{q}(\mathbf{a})$ is defined via the Rodrigue parametrization, i.e.,
    \begin{align}
    \delta\mathbf{q}(\mathbf{a}) = \frac{1}{\sqrt{4 + |\mathbf{a}|^2}}
    \begin{bmatrix}
    \mathbf{a} \\ 2
    \end{bmatrix},
    \end{align}
    where $\mathbf{a} \in \mathbb{R}^3$.
    
    The central idea of this approach is to treat the deviation vector, $\mathbf{a}$, as the latent variable and to estimate it by means of nonlinear Bayesian filtering.  
    To this end, a process model for the orientation deviation is needed. 
    In \cite{markley2004attitude}, it has been shown that the continuous time dynamics of the deviation vector can be written as 
    \begin{align}
    \dot{\mathbf{a}} = \left(I_3 + \frac{1}{4} \mathbf{a} \mathbf{a}^\top + \frac{1}{2} [\mathbf{a} \times]\right) \boldsymbol{\omega}\ ,
    \end{align}
    where $\boldsymbol{\omega} \triangleq [\omega_x\ \omega_y\ \omega_z]^\top$ represents the angular rate of the local frame with respect to the global frame. 
    We assume that $\mathbf{a}$ takes small values and approximate the expression as
    \begin{align} \label{eq:CTDevDyn}
    \dot{\mathbf{a}} \approx \left(I_3 + \frac{1}{2} [\mathbf{a} \times]\right) \boldsymbol{\omega}\ . 
    \end{align}
    
    Based on \eqref{eq:CTDevDyn}, a constant velocity model can easily be specified in continuous time, 
    \begin{align} \label{eq:CTRotDyn}
    \dot{\begin{bmatrix}
    	\mathbf{a} \\
    	\boldsymbol{\omega}
    	\end{bmatrix}}
    = 
    \begin{bmatrix}
    \left(I_3 + \frac{1}{2} [\mathbf{a} \times]\right) \boldsymbol{\omega} \\
    0_{3\times 1}
    \end{bmatrix} + 
    \begin{bmatrix}
    0_3 \\ 
    I_3
    \end{bmatrix}
    \boldsymbol{\alpha}\ ,
    \end{align}
    where $\boldsymbol{\alpha}$ denotes the rotational acceleration vector, and it is modeled as zero-mean white Gaussian noise with covariance ${\text{cov}[\boldsymbol{\alpha}(t),\boldsymbol{\alpha}(t')] = \delta(t-t') \Sigma_\alpha}$, where ${\Sigma_\alpha = \sigma_\alpha^2 I_3}$. 
    
    We need to discretize this system to be able to plug it in \eqref{eq:DTStateMatrices}. 
    Notice that \eqref{eq:CTRotDyn} is a nonlinear dynamic model. 
    Therefore, we will first linearize the equation around some point of interest and then discretize the resulting linearized model. 
    Let us first rewrite \eqref{eq:CTRotDyn} as 
    \begin{equation} \nonumber
    {\dot{\mathbf{x}}^r = f({\mathbf{x}^r}) + B \boldsymbol{\alpha}}\ ,
    \end{equation}
    where $\mathbf{x}^r \triangleq \left[\mathbf{a}^\top \ \boldsymbol{\omega}^\top \right]^\top$, and $f(\cdot)$ and $B$ are determined accordingly. 
    To linearize the model, we will substitute  $f(\cdot)$ by its first order Taylor series approximation. 
    \begin{subequations} \label{eq:TaylorApp1}
    	\begin{align}
    	f(\mathbf{x}^r) \approx f(\hat{\mathbf{x}}^r) + A_k^r (\mathbf{x}^r - \hat{\mathbf{x}}^r),\ 
    	\end{align}
    	where
    	\begin{align}
    	A_k^r = \frac{d}{d\mathbf{x}^r} f(\mathbf{x}^r) |_{\mathbf{x}^r=\hat{\mathbf{x}}_{k|k}^r}\ . 
    	\end{align}
    \end{subequations}
    The linearization is performed around the best available point estimate which is the mean of the previous posterior, i.e.,  ${\hat{\mathbf{x}}_{k|k}^r = \left[\hat{\mathbf{a}}_{k|k}^\top \ \hat{\boldsymbol{\omega}}_{k|k}^\top \right]^\top}$. 
    Note that after each measurement update of the filter, the reference orientation is updated by the estimated orientation deviation using the quaternion product, and then the deviation vector is reset to zero (see Appendix \ref{sec:App_orient} for the details). 
    Therefore, $\hat{\mathbf{a}}_{k|k}$ is equal to the zero vector.  
    Consequently, the Taylor series approximation in \eqref{eq:TaylorApp1} reduces to
    \begin{subequations}
    	\begin{align}
    	f(\mathbf{x}^r) = A_k^r \mathbf{x}^r,
    	\end{align}
    	where
    	\begin{align}
    	A_k^r = 
    	\begin{bmatrix}
    	\frac{1}{2}[-\hat{\boldsymbol{\omega}}_{k|k}\times] & I_3\\
    	0_3 & 0_3
    	\end{bmatrix}.
    	\end{align}
    \end{subequations}
    
    The resulting linearized system is given by
    \begin{align}
    \dot{\mathbf{x}}^r = A_k^r\ \mathbf{x}^r + B \boldsymbol{\alpha}. 
    \end{align}
    
    Thereafter, we discretize this equation and end up with the following linear Gaussian model which expresses the dynamics of the rotational subsystem. 
    \begin{subequations} \label{eq:DTRotDyn}
    	\begin{align}
    	\mathbf{x}^r_{k+1} = F^r_k  \mathbf{x}^r_k + \mathbf{w}^r_k,\quad \mathbf{w}_k^r \sim \mathcal{N}(\mathbf{0}, Q_k^r),
    	\end{align}
    	where
    	\begin{align}
    	F^r_k &= \exp(A_k^r T)\ , \\ 
    	Q_k^r &= G_k \Sigma_\alpha G_k^\top\ , \\
    	G_k &= \left(\int_{0}^{T} \exp(A_k^r \tau) d\tau \right) B\ .
    	\end{align}
    \end{subequations}
    $\mathbf{x}^r_k \triangleq \left[\mathbf{a}_k^\top \ \boldsymbol{\omega}_k^\top \right]^\top$ is the rotational kinematic state, $F^r_k$ is the system matrix, $Q_k^r$  is the process noise covariance matrix, and $T$ is the sampling time. 
    $F^r_k$ and $Q_k^r$ are explicitly indicated to be time-varying as they are recalculated considering the new linearization point at each iteration of the filter.
    For the complete details of the matrices in \eqref{eq:DTRotDyn}, see Appendix \ref{sec:App_RotDyn}. 
    
    The derivation of the given discrete-time state space model follows a standard approach, see, for example, \cite[Ch. 12.2]{gustafsson2010statistical}. 
    It is well known that the fidelity of the resulting model depends on the accuracy of the rotational state estimate as $F^r_k$ and $Q_k^r$ are approximately computed regarding the local linearization in \eqref{eq:TaylorApp1}. 

    \subsection{Measurement Model}
    In this subsection, the measurement model expressing the relation between the measurements and the state variables is derived to complete the state space model. 
    In general, we assume that there are multiple point measurements returned from an object at time $k$, which can be represented by the set $\{\mathbf{m}_{k,l}\}^{n_k}_{l=1}$. 
    A single measurement can be expressed as 
    \begin{align} \label{eq:measModelLocal}
    \mathbf{m}_{k,l} = \mathbf{c}_k + \mathbf{p}_{k,l}\ f((\theta, \phi)_{k,l}) + \bar{\mathbf{e}}_{k,l},
    \  \bar{\mathbf{e}}_{k,l} \sim \mathcal{N}(\mathbf{0},\bar{R}).
    \end{align}
    $\mathbf{c}_k$ is the center of the object at time $k$; 
    $(\theta, \phi)_{k,l}$ is the spherical angle pair indicating the measurement source on the object surface that originates $\mathbf{m}_{k,l}$; 
    $\mathbf{p}_{k,l}$ is the unit-length vector that starts from the object center and points towards the measurement source; 
    $f(\cdot)$ is the radial function; 
    and $\bar{\mathbf{e}}_{k,l}$ stands for the zero-mean Gaussian measurement noise with covariance $\bar{R}$. 
    
    Notice that for the measurement $\mathbf{m}_{k,l}$ in \eqref{eq:measModelLocal}, the underlying measurement source is unknown, and hence the corresponding $\mathbf{p}_{k,l}$ and $(\theta, \phi)_{k,l}$ are not available. 
    As an approximate approach, we hereby express these variables by utilizing the kinematic variables and the measurement. 
    
    Firstly, $\mathbf{p}_{k,l} (\mathbf{c}_k, \mathbf{m}_{k,l})$ is defined to be the unit-length vector starting from the object center and pointing to the measurement as 
    \begin{equation} 
    \mathbf{p}_{k,l} (\mathbf{c}_k, \mathbf{m}_{k,l}) = \frac{\mathbf{m}_{k,l} - \mathbf{c}_k}{\| \mathbf{m}_{k,l} - \mathbf{c}_k \|}.
    \end{equation}
    
    Then, for $(\theta, \phi)_{k,l}$ we need an intermediate representation of $\mathbf{m}_{k,l}$ by resolving it in the local coordinate frame. 
    It is simply obtained by the successive transformations of translation and rotation as 
    \begin{equation} \label{eq:measTransformation}
    \mathbf{m}_{k,l}^L \left(\mathbf{c}_k, \mathbf{q}_k, \mathbf{m}_{k,l} \right)= \underbrace{R_G^L(\mathbf{q}_k)}_{Rotation} \  \underbrace{(\mathbf{m}_{k,l} - \mathbf{c}_k)}_{Translation}. 
    \end{equation}
    $R_G^L(\mathbf{q})$ is the rotation matrix from the global to the local frame. 
    
    The relation between $\mathbf{m}_{k,l}$ and $\mathbf{m}_{k,l}^L$  is illustrated in Fig. \ref{fig:kumru2}. 
    $\mathbf{m}_{k,l}^L$ can be interpreted as a phantom measurement in the local frame, and it will only be exploited to find out the spherical angle pair, ${(\theta_{k,l}, \phi_{k,l})}$, associated to $\mathbf{m}_{k,l}$. 
    The figure depicts the object by a spherical shape to provide a straightforward description although the procedure applies to any arbitrary object. 
    Then, ${(\theta_{k,l}, \phi_{k,l})}$ is easily computed by converting $\mathbf{m}_{k,l}^L$ into the spherical coordinates by
    \begin{subequations} \label{eq:sphericalAngleLocalMeas}
    	\begin{align}
    	\theta_{k,l} &= \arctan\left(y^L/ x^L\right), \\
    	\phi_{k,l} &= \arctan \left(z^L/\sqrt{\left(x^L\right)^2 + \left(y^L\right)^2}\right),
    	\end{align}
    \end{subequations}
    where $\mathbf{m}_{k,l}^L \triangleq (x^L, y^L, z^L)$.
    
    Subsequently, the measurement equation in \eqref{eq:measModelLocal} is rewritten by 
    \begin{align} \label{eq:measModel}
    \mathbf{m}_{k,l} = \mathbf{c}_k + \mathbf{p}_{k,l} (\mathbf{c}_k, \mathbf{m}_{k,l}) f\left( \boldsymbol{\gamma}_{k,l}(\mathbf{c}_k, \mathbf{q}_k, \mathbf{m}_{k,l}) \right) + \bar{\mathbf{e}}_{k,l}\ ,
    \end{align}
    where the spherical angle pair is indicated by ${\boldsymbol{\gamma}_{k,l} \triangleq (\theta_{k,l}, \phi_{k,l})}$ for brevity. 
    Note that ${\boldsymbol{\gamma}_{k,l}}$ is a function of $\mathbf{c}_k$, $\mathbf{q}_k$ and $\mathbf{m}_{k,l}$ as implied by \eqref{eq:measTransformation} and \eqref{eq:sphericalAngleLocalMeas}. 
    
    Finally, the GP representation for the radial function given in \eqref{eq:extentStateSpace} is substituted, and we end up with the following measurement model. 
    \begin{subequations} 
    	\begin{align} \label{eq:gpeotMeasModel}
    	\mathbf{m}_{k,l} &= \mathbf{c}_k + \mathbf{p}_{k,l} \left[ H^\mathbf{f}\left( \boldsymbol{\gamma}_{k,l}(\mathbf{c}_k, \mathbf{q}_k, \mathbf{m}_{k,l}) \right) \mathbf{f}_k + e^\mathbf{f}_{k,l} \right]
    	+ \bar{\mathbf{e}}_{k,l} \nonumber \\
    	&= \underbrace{\mathbf{c}_k + \tilde{H}\left( \mathbf{c}_k, \mathbf{q}_k, \mathbf{m}_{k,l} \right) \mathbf{f}_k}_{=\tilde{\mathbf{h}}(\mathbf{x}_k, \mathbf{m}_{k,l})} 
    	+ \underbrace{\mathbf{p}_{k,l}\ e^\mathbf{f}_{k,l} + \bar{\mathbf{e}}_{k,l}}_{=\mathbf{e}_{k,l}} \nonumber  \\
    	&= \tilde{\mathbf{h}}(\mathbf{x}_k, \mathbf{m}_{k,l}) + \mathbf{e}_{k,l}, \quad \mathbf{e}_{k,l} \sim \mathcal{N} (\mathbf{0}, R_{k,l}), 
    	\end{align}
    	where
    	\begin{align}
    	\tilde{H}\left( \mathbf{c}_k, \mathbf{q}_k, \mathbf{m}_{k,l}\right) &= \mathbf{p}_{k,l}\ H^\mathbf{f}\left( \boldsymbol{\gamma}_{k,l}(\mathbf{c}_k, \mathbf{q}_k, \mathbf{m}_{k,l})\right),\\
    	R_{k,l} &= \mathbf{p}_{k,l}\ R_{k,l}^\mathbf{f}\  \mathbf{p}_{k,l}^\top + \bar{R}\ ,\\
    	\mathbf{p}_{k,l} &= \mathbf{p}_{k,l} (\mathbf{c}_k, \mathbf{m}_{k,l})\ , \\
    	R_{k,l}^\mathbf{f} &= R^\mathbf{f} \left(\boldsymbol{\gamma}_{k,l}(\mathbf{c}_k, \mathbf{q}_k, \mathbf{m}_{k,l})\right).
    	\end{align}
    \end{subequations}	
    Notice that the additive noise term $\mathbf{e}_{k,l}$ in \eqref{eq:gpeotMeasModel} actually depends on the unknown state vector, and it is not necessarily Gaussian. 
    However, we deliberately ignore this dependence and assume a Gaussian density to form an approximate model that is appropriate for the employed inference scheme.
    
    \eqref{eq:gpeotMeasModel} implies an implicit measurement model as $\mathbf{m}_{k,l}$ can not be explicitly written as a function of the state vector and the measurement noise. 
    We collect all the terms on one side of the equation as 
    \begin{align} \label{eq:gpeotMeasModelImp}
    \mathbf{0} &= \underbrace{-\mathbf{m}_{k,l} + \tilde{\mathbf{h}}(\mathbf{x}_k, \mathbf{m}_{k,l})}_{=\mathbf{h}(\mathbf{x}_k, \mathbf{m}_{k,l})} + \mathbf{e}_{k,l} \nonumber\\
    &= \mathbf{h}(\mathbf{x}_k, \mathbf{m}_{k,l}) + \mathbf{e}_{k,l}, \quad \mathbf{e}_{k,l} \sim \mathcal{N} (\mathbf{0}, R_{k,l}).
    \end{align}
    
    The zero vector in \eqref{eq:gpeotMeasModelImp} can be interpreted as a pseudo-measurement, which is a nonlinear function of the state and the measurement, and it is corrupted by some additive Gaussian noise. 
    Implicit measurement models have been commonly utilized for different applications, which perform inference by various means of Kalman filtering (see, for example, \cite{baum2011shape}, \cite{dantanarayana2015extended}).

    \section{Inference} \label{sec:Inference}
    Having developed the state space model, the last step is to design an effective inference method to realize object tracking using point cloud measurements. 
    While there are various standard techniques to recursively compute the posterior distribution of the state vector, we employ an extended Kalman filter (EKF) due to the nonlinearities in the measurement model. 
    
    To be able to process multiple measurements $\{\mathbf{m}_{k,l}\}^{n_k}_{l=1}$ in a single recursion at time $k$, we first need to slightly modify the state space model. 
    To this end, the following measurement vector is created by concatenating the measurements together, 
    \begin{equation}
    \mathbf{m}_{k} = \left[ \mathbf{m}^\top_{k,1},\ \dots,\ \mathbf{m}^\top_{k,n_k} \right] ^ \top.	
    \end{equation}
    Then, the corresponding measurement equation can be simply written as
    \begin{subequations}
    	\begin{align}
    	\mathbf{0} &= \mathbf{h}(\mathbf{x}_k, \mathbf{m}_k) + \mathbf{e}_{k}, \quad \mathbf{e}_{k} \sim \mathcal{N} (\mathbf{0}, R_{k}),  \\
    	\mathbf{h}(\mathbf{x}_k, \mathbf{m}_k) &= \left[ \mathbf{h}(\mathbf{x}_k, \mathbf{m}_{k,1})^\top,\ \dots,\ \mathbf{h}(\mathbf{x}_k, \mathbf{m}_{k,n_k})^\top \right]^\top, \\
    	R_k &= \text{diag} \left[ R_{k,1},\ \dots,\ R_{k,n_k} \right]. 	
    	\end{align}
    \end{subequations}
    Notice that $R_k$ is formed as a block diagonal matrix by considering that the noise coupled to the individual measurements are mutually independent. 
    
    Consequently, the state space model considering the complete set of measurements reads as 
    \begin{subequations} \label{eq:SSModelSummary}
    	\begin{align}
    	\mathbf{x}_{k+1} &= F_k \mathbf{x}_{k} + \mathbf{w}_k,\quad \mathbf{w}_k \sim \mathcal{N}(\mathbf{0}, Q_k), \\
    	\mathbf{0} &= \mathbf{h}(\mathbf{x}_k, \mathbf{m}_k) + \mathbf{e}_{k}, \quad \mathbf{e}_{k} \sim \mathcal{N} (\mathbf{0}, R_{k}),  \\
    	\mathbf{x}_0 &\sim \mathcal{N}(\boldsymbol{\mu}_0, P_0). 
    	\end{align}
    \end{subequations}
    The EKF regards the above representation to recursively compute the estimate of the state vector, $\hat{\mathbf{x}}_{k|k}$. 
    Note that the gradient of the measurement function  $\frac{\partial \mathbf{h}_{k}(\mathbf{x}_k)}{\partial \mathbf{x}_k} $ can be derived analytically which is to be utilized in the measurement update phase of each recursion.
    
    For very large point clouds, one may want to optimize the computational characteristics of the filter, which is mainly determined by the inversion operation of the innovation covariance matrix (see Appendix B). 
    In this regard, a sequential update of the measurements can be preferred instead of a batch update. 
    The update can also be performed in the information form of the EKF, \cite[Ch. 3.5]{thrun2002probabilistic}.
	
    \section{3D Extent Tracking Using Projections} \label{sec:Proj}
    In the first part of this study, we developed a tracking algorithm which is essentially based on the radial function representation, $f(\theta, \phi)$, of the underlying 3D object shape. 
    This function is further approximated via some basis points at which the shape information is accumulated during inference. 
    Notice that as there are two input arguments of the radial function, the basis points are required to cover a two-dimensional space at a sufficient density to be able to capture the characteristics of the object shape. 
    Also note that the computational load and the memory storage scale with the number of basis points since they are included in the state vector and updated at each recursion. 
    A naive attempt to utilize fewer basis points for more efficient implementation will naturally result in a degraded representational power, potentially missing salient features of the 3D extent which might in turn deteriorate tracking accuracy. 
    
    In this section, we will seek for an alternative algorithm with improved computational properties. 
    This second approach essentially retains the basic structure of the previous one; however, it fundamentally differs in the description of the object shape. 
    In particular, multiple projections of the object are exploited to express the original 3D extent. 
    Accordingly, the problem is reformulated as tracking the object while simultaneously learning the contours of its projections. 
    This will eventually enable us to radically lower the number of basis points without compromising the representational power. 
    The next section presents the alternative extent model in details. 
    
    \subsection{Projection Model}
\begin{figure}[tb]
    	\centering
    	\includegraphics{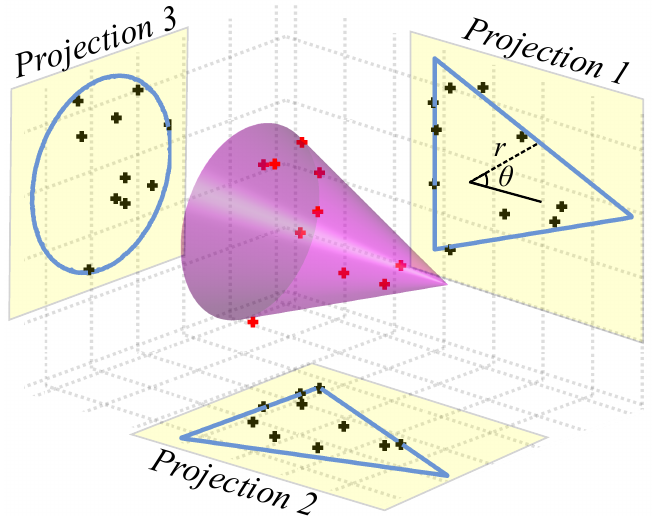}
    	\caption{Illustration of a cone-shaped object and the corresponding projection contours on three orthogonal planes. 
    		Point cloud measurements and their projections are shown by red and black plus signs, respectively. }
    	\centering
    	\label{fig:kumru3}
    \end{figure}

    It is a long-standing idea to exploit projections, silhouettes or images for expressing the corresponding 3D shape, \cite{rivers20103d, niem1994robust}. 
    Being inspired by these methods, we suggest to model the object extent using projections onto several planes. 
    Fig. \ref{fig:kumru3} illustrates the idea for an example object with cone shape. 
    In this case, the object is projected onto three orthogonal planes and the contours of these projections are essentially utilized to represent the original 3D shape. 
    In this exposition, we assume that three orthogonal projections can sufficiently approximate the 3D shape; however, the number of projections can be increased to be able to generalize to a broader class of objects. 
    For a systematic discussion on the objects which are exactly reconstructable from projections and the minimum number of projections necessary for reconstructing such objects, interested readers can refer to \cite{laurentini1997many}. 
    
    The contour of each projection can be described by a radial function in polar coordinates, i.e., $r = f(\theta)$, as shown in Fig. \ref{fig:kumru3}. 
    The radial function maps the polar angle, $\theta$, to the radial distance, $r$, between the projection center and the contour. 
    Notice that having only one input argument, this function can possibly be approximated by a less number of basis points leading to a tracking algorithm demanding less computational sources. 
     
    The rest of the derivation closely follows the first algorithm. 
    The unknown radial function on each projection plane is modeled by a GP, i.e., ${f(\theta) \sim \mathcal{GP}(\mu (\theta), k(\theta, \theta'))}$, whose mean function is taken to be constant $\mu (\theta) = \mu_r$, and the covariance function is defined as  
    \begin{equation} \label{eq:covFuncProj}
    k(\theta, \theta') = \sigma_f^2 e^{-\frac{2 \text{sin}^2\left(\frac{\theta-\theta'}{2}\right)}{l^2}} + \sigma_r^2. 
    \end{equation}
    Notice that the $\text{sin}(\cdot)$ term in \eqref{eq:covFuncProj} is used to induce a periodic covariance function, which in turn assures the periodicity of the radial function $f(\cdot)$ described by the given GP model, \cite{wahlstrom2015extended}.

    \textit{\textbf{Further Discussion:}}
    Expressing the 3D shape in terms of a collection of projection contours enables us to introduce separate probabilistic models for each contour to account for application-specific knowledge about the objects. 
    For example, many targets in driving environments, such as cars, vans and bicycles, possess a common characteristic in their projections onto the ground plane. 
    In particular, the radial function describing the corresponding projection contour appears to be periodic with $\pi$. 
    In this case, to encode this information into the GP model, the covariance function can be designed as 
    \begin{equation} \label{eq:covFuncSym}
    k\left(\theta, \theta^{\prime}\right) = \sigma_{f}^{2} e^{-\frac{\sin ^{2}\left(\theta-\theta^{\prime}\right)}{2 l^{2}}} + \sigma_r^2. 
    \end{equation}
    As the covariance function is periodic with $\pi$, the learned contours will comply with the actual characteristics of the projections as intended. 
    
    \subsection{State Space Model}
    In this subsection, the state space model relying on the extent description obtained by projection contours is to be constructed. 
    The state vector is defined as ${\mathbf{x}_{k} \triangleq \left[\bar{\mathbf{x}}_{k}^\top \ \mathbf{f}_k^\top \right]^\top}$ where ${\bar{\mathbf{x}}_{k}}$ includes the object kinematics and the extent is indicated by ${\mathbf{f}_k \triangleq \left[ {\mathbf{f}_k^1}^\top \ {\mathbf{f}_k^2}^\top \ {\mathbf{f}_k^3}^\top \right]^\top}$ as a collection of the projection contours. 
    More specifically, $\mathbf{f}^j_k$ is the parameterized description of the GP model for the radial function specifying the contour of the projection on the $j^{th}$ plane. 
    $\mathbf{c}_k$ is the center of the 3D object and $\mathbf{v}_k$ stands for the velocity of the center; 
    $\mathbf{q}_k$ is the unit quaternion vector. 

    \subsubsection{Measurement Model}
    The measurement model makes use of the local and global coordinate frames as defined earlier. 
    While the object motion is tracked in the global frame, the shape is described in the local frame. 
    In particular, the projection planes are fixed to the local frame so that the projections of the object onto the planes are kept unchanged at any time. 
    This allows to accumulate extent information over these planes by learning the latent contours of the projections. 
    
    Consider the 3D point cloud measurements acquired at time $k$, $\{\mathbf{m}_{k,i}\}^{n_k}_{i=1}$.
    Firstly, each measurement $\mathbf{m}_{k,l}$ is transformed into the local frame to obtain $\mathbf{m}_{k,l}^L$ by \eqref{eq:measTransformation}. 
    Thereafter, these local measurements are to be projected onto each plane to establish a relation between the projected measurements and the projection contour on the corresponding plane. 
    As an example, let $\mathbf{m}_{k,l}^L$ be projected onto the $j^{th}$ plane by 
    \begin{align} \label{eq:projMeas}
    \mathbf{m}^{j}_{k,l} &= P_j\ \mathbf{m}_{k,l}^L\ , 
    \end{align}  
    where $\mathbf{m}^j_{k,l} \in \mathbb{R}^2$ denotes the projection of $\mathbf{m}_{k,l}^L$, and ${P_j \in \mathbb{R}^{2x3}}$ is the projection matrix.  
    
    By assuming that the projections of the measurements are originated from a star convex extent model \cite{wahlstrom2015extended}, \cite{baum2014extended}, the projected measurement can be described as
    \begin{equation} \label{eq:projMeasInitial}
    \begin{split}
    \mathbf{m}^j_{k,l} = \mathbf{p}_{k,l}(\mathbf{c}_k, \mathbf{q}_k, \mathbf{m}_{k,l}) f^j(\theta_{k,l}(\mathbf{c}_k, &\mathbf{q}_k, \mathbf{m}_{k,l})) + \bar{\mathbf{e}}_{k,l}, \\
    {}&\bar{\mathbf{e}}_{k,l} \sim \mathcal{N}(0,\bar{R}), 
    \end{split}
    \end{equation}
    where $f^j(\cdot)$ is the radial function expressing the contour of the projection on the $j^{th}$ plane; 
    $\mathbf{p}_{k,l}(\mathbf{c}_k, \mathbf{q}_k, \mathbf{m}_{k,l})$ is the unit-length vector pointing from the projection center towards the measurement; 
    and $\bar{\mathbf{e}}_{k,l}$ is the Gaussian measurement noise with covariance matrix $\bar{R}$. 
    Notice that unlike \eqref{eq:measModel}, the center position is not superposed in \eqref{eq:projMeasInitial} as the projection is specified to be centered at the origin of the corresponding plane. 
    
    In \eqref{eq:projMeasInitial}, the underlying source of $\mathbf{m}^j_{k,l}$ is actually unknown. 
    Therefore, we resort to an approximate approach and formulate the expressions on the right hand side of the measurement model as functions of the kinematic variables and the measurement $\mathbf{m}_{k,l}$ itself.
    
    The polar angle $\theta_{k,l}$ associated with the projected measurement can be computed as
    \begin{align}
    \theta_{k,l} (\mathbf{c}_k, \mathbf{q}_k, \mathbf{m}_{k,l}) = \angle{\mathbf{m}^j_{k,l}}\ .
    \end{align}
    Besides, the unit-length vector $\mathbf{p}_{k,l}$ is obtained by 
    \begin{align} 
    \mathbf{p}_{k,l} (\mathbf{c}_k, \mathbf{q}_k, \mathbf{m}_{k,l}) &= \frac{\mathbf{m}^j_{k,l}}{\| \mathbf{m}^j_{k,l} \|}\ .
    \end{align}
    
    The next step is to plug the GP representation for the radial function into \eqref{eq:projMeasInitial} as  
    \begin{subequations}
    	\begin{align} \label{eq:measModelIntProj}
    	\mathbf{m}^j_{k,l} = \tilde{H}(\mathbf{c}_k, \mathbf{q}_k, \mathbf{m}_{k,l})\ \mathbf{f}^j_k + \tilde{\mathbf{e}}_{k,l}, \ 
    	\tilde{\mathbf{e}}_{k,l} \sim \mathcal{N}(0, \tilde{R}_{k,l}),
    	\end{align} 
    	where
    	\begin{align}
    	\tilde{H}(\mathbf{c}_k, \mathbf{q}_k, \mathbf{m}_{k,l}) &= \mathbf{p}_{k,l} H^\mathbf{f}(\theta_{k,l}(\mathbf{c}_k, \mathbf{q}_k, \mathbf{m}_{k,l}))\ , \\
    	\tilde{\mathbf{e}}_{k,l} &= \mathbf{p}_{k,l} e^\mathbf{f}_{k,l} + \bar{\mathbf{e}}_{k,l}\ , \\
    	\tilde{R}_{k,l} &= \mathbf{p}_{k,l} R^\mathbf{f}_{k,l} \mathbf{p}_{k,l}^\top + \bar{R}\ , \\
    	\mathbf{p}_{k,l} &= \mathbf{p}_{k,l} (\mathbf{c}_k, \mathbf{q}_k, \mathbf{m}_{k,l})\ , \\
    	R_{k,l}^\mathbf{f} &= R^\mathbf{f} \left(\theta_{k,l}(\mathbf{c}_k, \mathbf{q}_k, \mathbf{m}_{k,l})\right)\ .
    	\end{align}
    \end{subequations}
    As before, the additive noise term $\tilde{\mathbf{e}}_{k,l}$ in \eqref{eq:measModelIntProj} is actually a function of the unknown state vector, and it is not necessarily Gaussian. 
    Again, we ignore this dependence and assume a Gaussian density so that the resulting approximate model lets us use an EKF for inference.
    
    The projected measurements are not necessarily located on the contour, instead some of them may fall within the interior of the projection area as depicted in Fig. \ref{fig:kumru3}. 
    Accounting for this observation, the measurement model is modified as 
    \begin{align}
    \mathbf{m}^j_{k,l} &= s_{k,l} \tilde{H}(\mathbf{c}_k, \mathbf{q}_k, \mathbf{m}_{k,l})\ \mathbf{f}^j_k + \tilde{\mathbf{e}}_{k,l}\ , 
    \end{align}
    where $s_{k,l} \in [0,1]$ is a random scaling factor. 
    We approximate $s$ as a Gaussian random variable, i.e., $s_{k,l} \sim \mathcal{N} (\mu_s, \sigma^2_s)$, \cite{baum2009random}, \cite{wahlstrom2015extended}, \cite{baum2014extended}, since an EKF will be employed for inference. 
    In an ideal case, a particular angle dependent one-dimensional probability distribution for the random scaling factor should be chosen depending on the object-sensor geometry, object's shape and the projection planes.

    Considering the characteristics of the scaling factor, the measurement model can be rewritten as
    \begin{subequations}
    	\begin{align}
    	\mathbf{m}^j_{k,l} &= \underbrace{\mu_s  \tilde{H}_{k,l}\  \mathbf{f}^j_k}_{=\tilde{\mathbf{h}}^j(\mathbf{x}_k, \mathbf{m}_{k,l})} +\ \underbrace{(s_{k,l} - \mu_s) \tilde{H}_{k,l}\  \mathbf{f}^j_k + \tilde{\mathbf{e}}_{k,l}}_{=\mathbf{e}^j_{k,l}} \nonumber\\ 
    	&= \tilde{\mathbf{h}}^j(\mathbf{x}_k, \mathbf{m}_{k,l}) + \mathbf{e}^j_{k,l}, \quad \mathbf{e}^j_{k,l} \sim \mathcal{N} (0, R^j_{k,l})\ ,
    	\end{align}
    	where
    	\begin{align}
    	R^j_{k,l} &= \sigma_s^2 \tilde{H}_{k,l}\ \mathbf{f}^j_k\ {\mathbf{f}^j_k}^\top \tilde{H}_{k,l}^\top + \tilde{R}_{k,l}\ , \\
    	\tilde{H}_{k,l} &= \tilde{H}(\mathbf{c}_k, \mathbf{q}_k, \mathbf{m}_{k,l})\ .
    	\end{align}	
    \end{subequations}
    
    Then, the expression for the projected measurement is substituted into this equation as
    \begin{align} \label{eq:projectedMeasurement2}
    P_j R_G^L(\mathbf{q}_k) (\mathbf{m}_{k,l} - \mathbf{c}_k) = \tilde{\mathbf{h}}^j(\mathbf{x}_k, \mathbf{m}_{k,l}) + \mathbf{e}^j_{k,l}\ . 
    \end{align}
    
    Finally, collecting the terms on one side of the equation, we end up with the following implicit measurement model with additive Gaussian measurement noise.
    \begin{align} 
    \mathbf{0} &= \underbrace{-P_j R_G^L(\mathbf{q}_k) (\mathbf{m}_{k,l} - \mathbf{c}_k) + \tilde{\mathbf{h}}^j(\mathbf{x}_k, \mathbf{m}_{k,l})}_{=\mathbf{h}^j(\mathbf{x}_k, \mathbf{m}_{k,l})}  + \mathbf{e}^j_{k,l}\nonumber\\ 
    &= \mathbf{h}^j(\mathbf{x}_k, \mathbf{m}_{k,l}) + \mathbf{e}^j_{k,l}
    \end{align}

    This measurement model together with the process model introduced in Section \ref{sec:GPETT3DProcessModel} establishes the state space model. 
    \subsection{Inference}
    Similar to the former case, an EKF is employed to realize recursive inference. 
    To process all measurements instantaneously at the update phase of the filter, the complete measurement equation is written as 
    \begin{subequations}
    	\begin{align}
    	\mathbf{0} &= \mathbf{h}(\mathbf{x}_k, \mathbf{m}_{k}) + \mathbf{e}_k , \quad \mathbf{e}_{k} \sim \mathcal{N} (0, R_{k}), 
    	\end{align}
    	where
    	\begin{align}
    	\mathbf{h}(&\mathbf{x}_k, \mathbf{m}_{k}) =	\left[ {\mathbf{h}_{k}^1}^\top,\ {\mathbf{h}_{k}^2}^\top,\ {\mathbf{h}_{k}^3}^\top \right]^\top, \\
    	\mathbf{m}_{k} &= \left[ \mathbf{m}^\top_{k,1},\ \dots,\ \mathbf{m}^\top_{k,n_k} \right] ^ \top, \\
    	\mathbf{h}_{k}^j &= \mathbf{h}^j(\mathbf{x}_k, \mathbf{m}_{k}) = 
    	\left[ {\mathbf{h}^j}_{k,1}^\top,\ \dots, {\mathbf{h}^j}_{k,n_k}^\top \right]^\top,  \\
    	\mathbf{h}^j_{k,l} &= \mathbf{h}^j(\mathbf{x}_k, \mathbf{m}_{k, l})\ , \\
    	R_k &= \text{diag} \left[ R_k^1,\ R_k^2,\ R_k^3 \right], \\
    	R_k^j &= \text{diag} \left[ R_{k,1}^j,\ \dots, R_{k,n_k}^j \right]\ \text{for} \ j \in \{1,2,3\}. 
    	\end{align}
    \end{subequations}
	
    \section{Results} \label{sec:Results}
    In this section, the performance of the proposed algorithms is evaluated on both simulated and real measurements in Section \ref{sec:Result_Simulated} and \ref{sec:Result_RealData}, respectively. 
    To be able to present the results in a comparative manner, we also regard a standard random matrix-based extended object tracker \cite{feldmann2011tracking}, denoted as RM.
    Describing the extent by an ellipsoid, the RM model has proven to be extremely robust for a wide range of scenarios. 
    Therefore, it serves as a solid basis to assess the tracking performance of the suggested models. 
    Throughout this section, we will refer to the first proposed method as `GPEOT' (short for GP-based extended object tracker), while `GPEOT-P' will stand for the second approach considering the projections. 

    \begin{figure*} [t]
    	\centering
    	\subfloat[GPEOT]{\includegraphics[trim= 0 0 0 0,clip, width=0.322\textwidth]{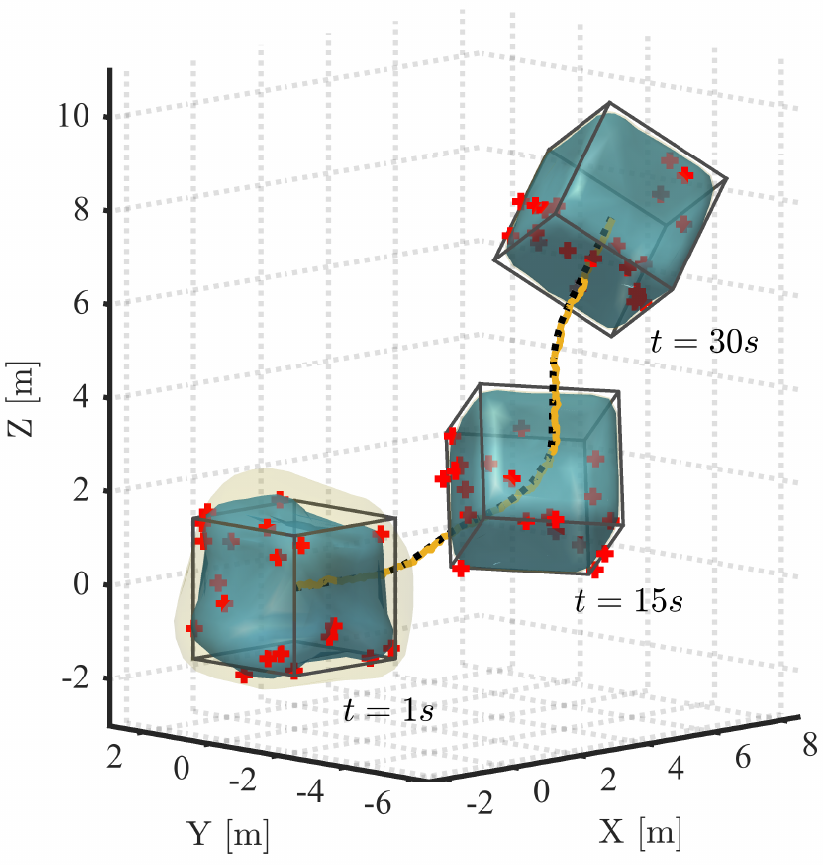}}
    	\subfloat[GPEOT-P]{\includegraphics[trim= 0 0 0 0,clip, width=0.322\textwidth]{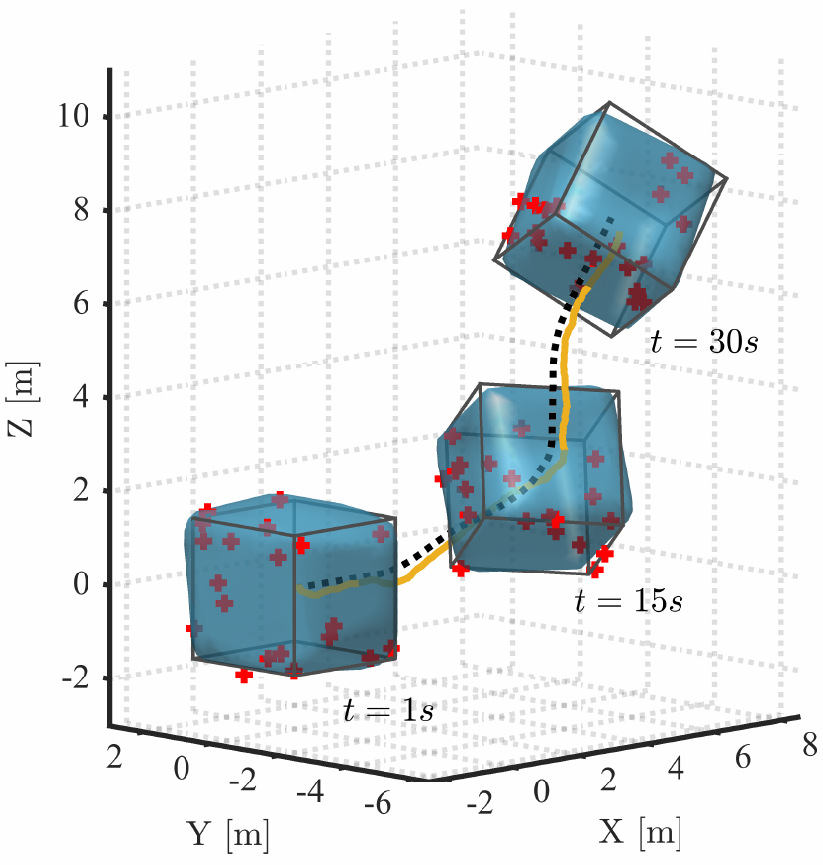}\label{fig:MatUTCone_gpettP}}
    	\subfloat[RM]{\includegraphics[trim= 0 0 0 0,clip, width=0.322\textwidth]{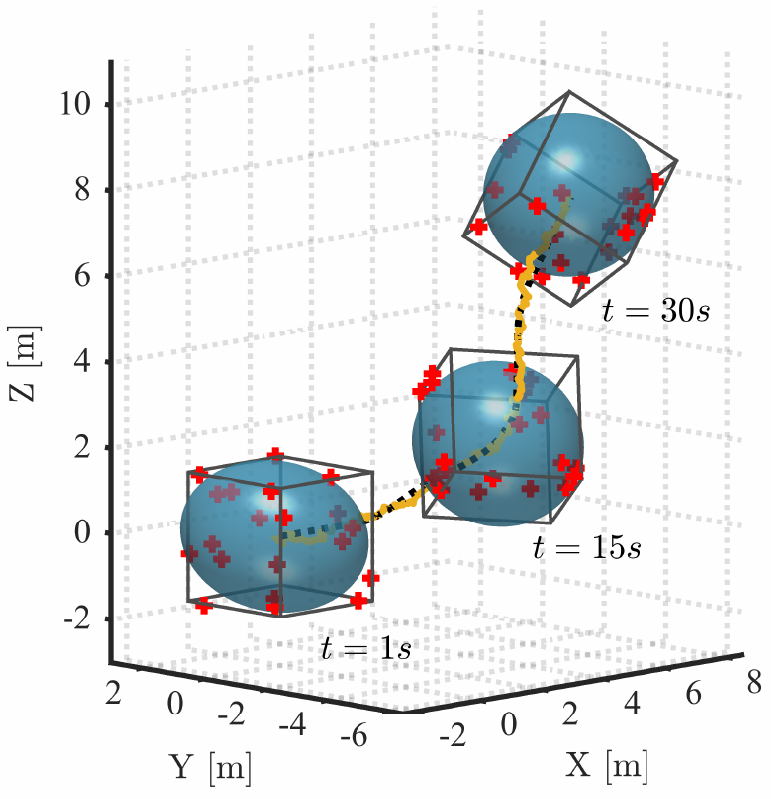}}
    	\caption{Typical results for the cube-shaped object during the complex maneuvering experiment. 
    		(Blue surface and black box visualize the estimated and the true extent of the object, respectively. 
    		In Fig. (a) yellow surface indicates the confidence interval of one standard deviation. 
    		Red plus signs are the point measurements. 
    		Solid yellow and dashed black curves are the estimated and true trajectory, respectively.)}
    	\label{fig:kumru4}
    \end{figure*}
    \begin{figure} [t]
    	\centering
    	\subfloat[Linear motion experiments]{\includegraphics[trim= 35 135 30 0,clip, width=\columnwidth]{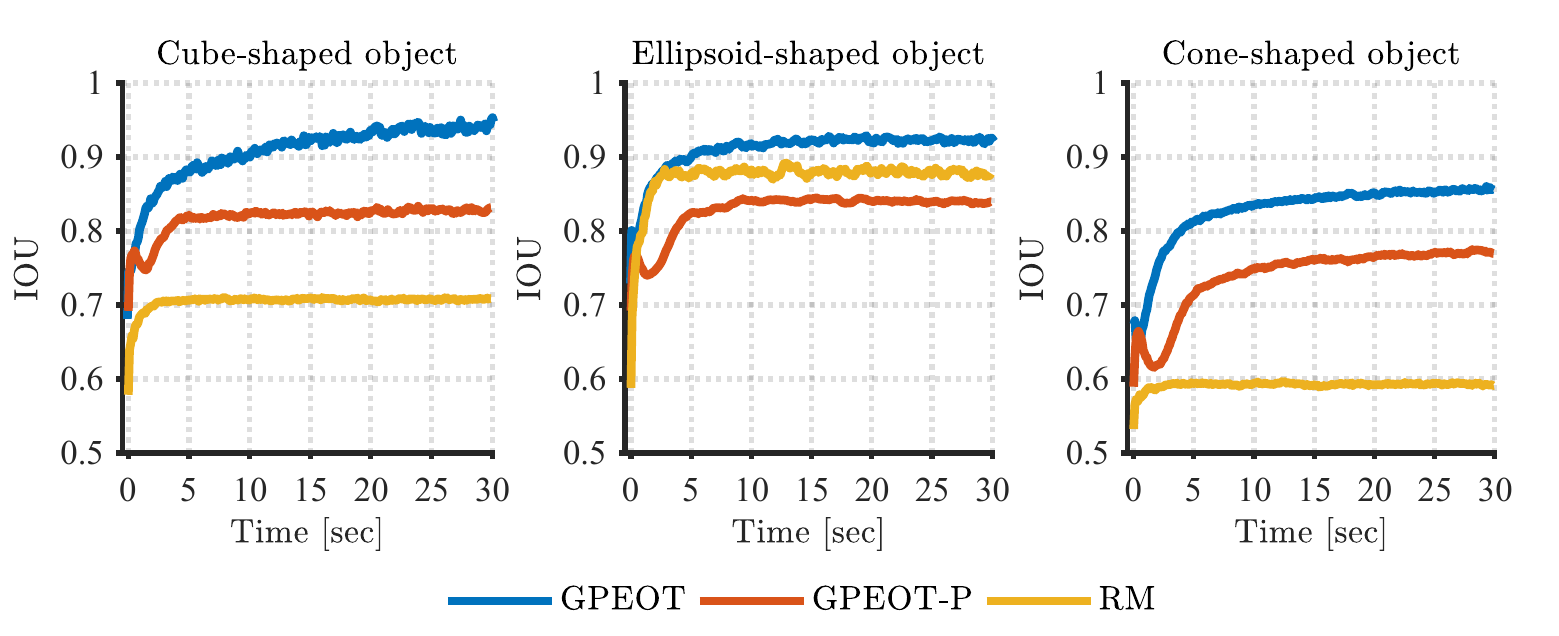}}\\
    	\subfloat[Complex maneuvering experiments]{\includegraphics[trim= 35 135 30 0,clip, width=\columnwidth]{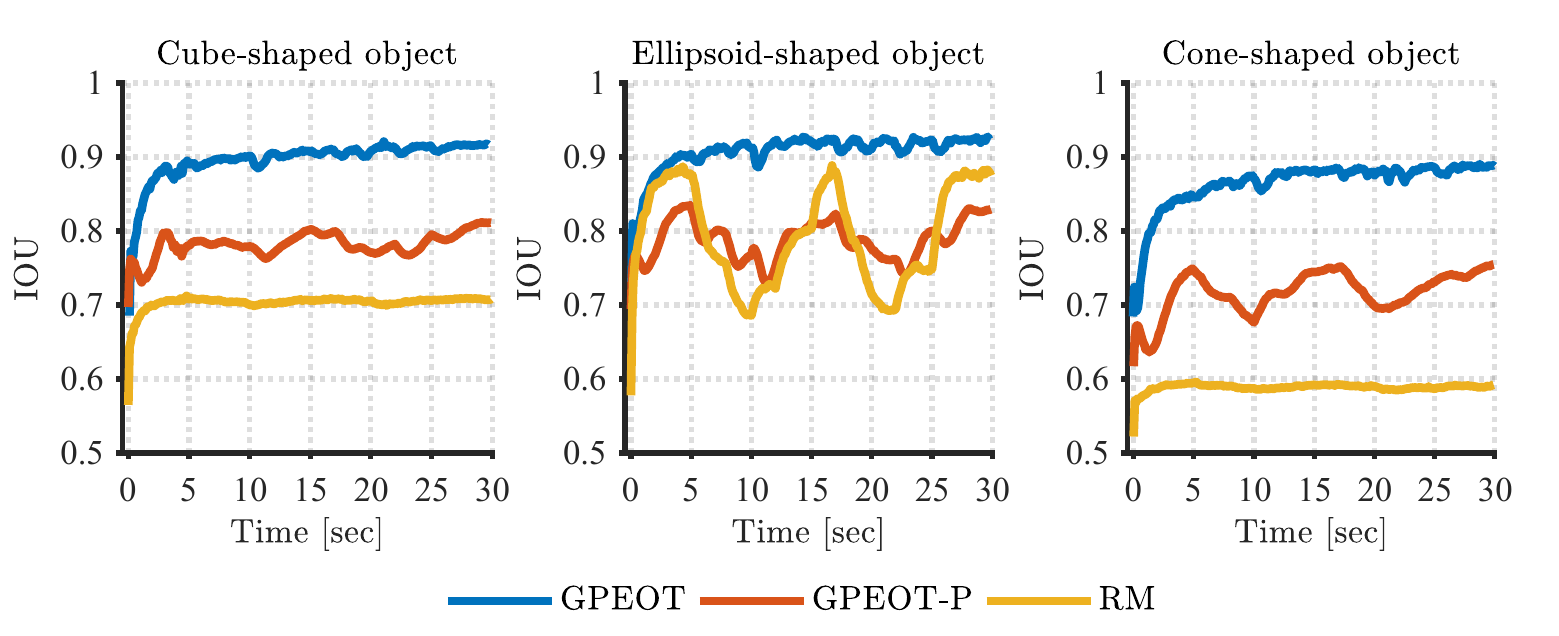}}
    	\caption{Intersection-Over-Union (IOU) plots. (The results are averaged over 100 MC runs.)}
    	\label{fig:kumru5}
    \end{figure}

    \subsection{Experiments with Simulated Measurements} \label{sec:Result_Simulated}
    
    To demonstrate the performance of the algorithms, various simulation experiments are conducted. 
    Section \ref{sec:Result_Matlab} examines the setting where the point cloud measurements are simulated in \MAT\textsuperscript{\textregistered} for several dynamic objects with basic shapes. 
    In this case, the measurements are randomly sampled from the objects' surfaces. 
    In Section \ref{sec:Result_Blensor}, we make use of a specialized sensor simulation environment, namely Blensor, \cite{gschwandtner2011blensor}, that generates measurements for realistic sensor and vehicle models. 
    
    \subsubsection{\MAT\ Simulations} \label{sec:Result_Matlab}
    In this subsection, the algorithms process point cloud measurements which are generated from random sources on the object surface and perturbed by additive Gaussian noise. 
    Three different-shaped objects, e.g., cube, ellipsoid and cone, are tracked during the experiments. 
    The dimensions of the objects are as follows: the length of the edge of the cube is 3 m, the semi-axes of the ellipsoid are (2.5, 1, 1) m in length, and the base radius and height of the cone are 1.5 m and 4 m, respectively. 
    
    The overall performance is evaluated based on the Intersection-Over-Union (IOU) measure given by
    \begin{equation}
    \text{IOU}(S_{true}, \hat{S}) = \frac{\text{volume}(S_{true}\cap\hat{S})}{\text{volume}(S_{true}\cup\hat{S})}, 
    \end{equation}
    where $S_{true}$ is the true object shape, and $\hat{S}$ stands for the estimate.  
    Notice that IOU simultaneously accounts for the quality of the estimates of the kinematics and the extent. 
    In other words, an algorithm needs to produce accurate tracking outputs together with precise shape description to attain high IOU scores. 
    Also note that in our discussion we deliberately exclude the RMSE measure for the position estimates since the suggested shape models do not imply a unique center definition; instead, different center positions with compatible radial functions can accurately represent the same object. 
    
    We also compute the root mean squared error (RMSE) of the object velocity, defined as
    \begin{align}
    	\text{RMSE}(\hat{\mathbf{v}}_{k}, \mathbf{v}_{k}) = \sqrt{\frac{1}{N}\sum_{k=1}^N |\hat{\mathbf{v}}_{k} - \mathbf{v}_{k}|^2}, 
    \end{align}
    where $\hat{\mathbf{v}}_{k}$ and $\mathbf{v}_{k}$ indicate the estimated and the true velocity vector at time $k$, respectively;
    $|\cdot|$ indicates the Euclidean norm.

    \begin{figure}[t]
    	\centering
    	\includegraphics{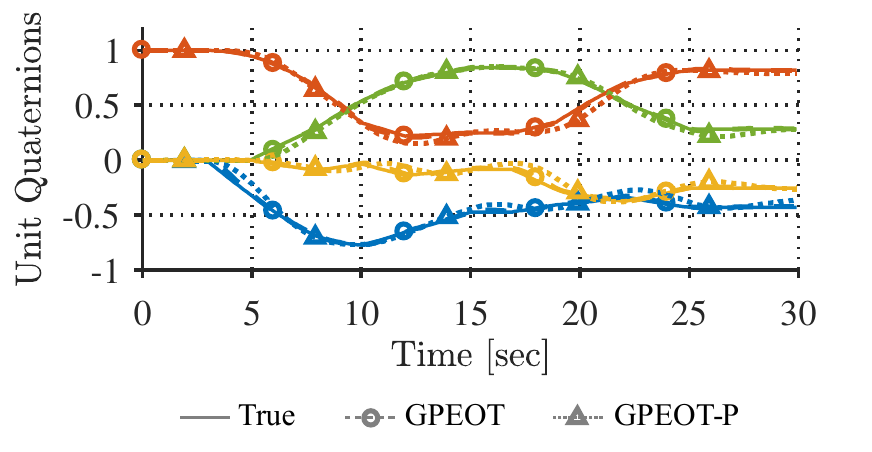}
    	\caption{True and estimated unit quaternions for the cube-shaped object during the complex maneuvering experiment. 
    		(The estimates are averaged over 100 MC runs. 
    		Color code is blue: $q_0$, green: $q_1$, yellow: $q_2$, red: $q_3$.)}
    	\centering
    	\label{fig:kumru6}
    \end{figure}
    
    \begin{figure}[t]
    	\centering
    	\includegraphics[trim= 14 0 0 0,clip]{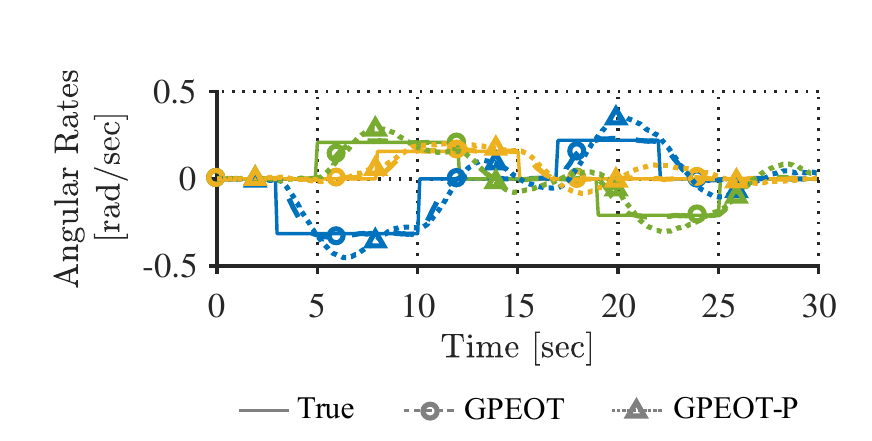}
    	\caption{True and estimated angular rates for the cube-shaped object during the complex maneuvering experiment. 
    		(The estimates are averaged over 100 MC runs. 
    		Color code is blue: $\omega_x$, green: $\omega_y$, yellow: $\omega_z$.)}
    	\label{fig:kumru7}
    \end{figure}
    
    Two different scenarios are studied in the simulations: a linear motion and a complex maneuver. 
    In the first case, objects move along a linear trajectory at a constant speed of 10 m/s. 
    In the second experiment, the object follows a curved path while performing combined rotations around different axes. 
    Throughout the trajectory, the linear speed is kept constant at 0.5 m/s. 
    At each instant, 20 point measurements are originated from random sources which are sampled from a uniform distribution defined over the object surface. 
    Each point measurement is perturbed by i.i.d. Gaussian noise with covariance $0.1^2 I_3$. 
    The measurements are produced at 10 Hz, hence the sampling time of all algorithms is set to $T = 0.1$. 
    We should note that the simulated measurements used in this section do not follow the actual characteristics of range scanners, which partially delineate the object's surface due to self-occlusions and generate sparse measurements with increased distance between the sensor and the object.
    We will investigate experiments regarding data collected by real sensors in Section \ref{sec:Result_RealData}. 
    
    For GPEOT, the process noise standard deviations are set to $\sigma_c=0.1$ and $\sigma_\alpha=0.1$, and 
    $\lambda=0.99$ is used for the extent dynamics; 
    the hyper-parameters of the GP model are set to $\mu_r = 0$, $\sigma_f = 1$, $\sigma_r = 0.2$, $l = \pi/8$; 
    $\bar{R} = 0.1^2 I_3$ is used for the measurement noise variance; 
    and the extent is represented by 642 basis points which are evenly spaced with respect to their spherical angles. 
    For GOEOT-P, the process noise standard deviations are set to $\sigma_c=0.1$ and $\sigma_\alpha=0.4$, and 
    $\lambda=0.99$ is used for the extent dynamics; 
    the hyper-parameters of the GP model are set to $\mu_r = 0$, $\sigma_f = 1$, $\sigma_r = 0.2$, $l = \pi/5$; $\bar{R} = 0.1^2 I_2$ is used for the projected measurement noise variance; 
    each projection contour is represented by 50 basis points which are equidistantly located in $[0, 2\pi]$; 
    and the parameters of the scaling factor are set to $\mu_s = \frac{5}{6}$ and $\sigma^2_s = \frac{1}{18}$. 
    The projection planes are selected to be the $xy$, $xz$ and $yz$ planes, hence the corresponding projection matrices are 
    ${P_1 = \big[\begin{smallmatrix}
    	1 & 0 & 0\\
    	0 & 1 & 0
    	\end{smallmatrix}\big],\  
    	P_2 = \big[\begin{smallmatrix}
    	1 & 0 & 0\\
    	0 & 0 & 1
    	\end{smallmatrix}\big],\ 
    	P_3 = \big[\begin{smallmatrix}
    	0 & 1 & 0\\
    	0 & 0 & 1
    	\end{smallmatrix}\big]}$. 
    All algorithms use the same prior distribution for the kinematics. 
    The prior distribution for the extent state is directly specified by the corresponding GP model. 
    
    The RM model assumes that the measurements have a normal spread, which is specified by the extent state, and hence they are not confined to be originated by the object's surface. 
    In this regard, the method can be considered to be relatively less suitable for sensors like LIDARs, where the measurements are generated exclusively from the surface.  
    The scaling factor included in the RM model is of practical value for such cases as it accounts for the discrepancy between the sensor's actual behavior and the mentioned assumption, \cite{feldmann2011tracking}. 
    To obtain a competent performance, we manually optimized the parameters of the RM model: The scaling factor and the extension time constant are set to 1/3 and 1, respectively. 
    
    Due to page limitations, we hereby present some instances of our findings as representative examples. 
    Typical results for the cube-shaped object performing the complex maneuver are illustrated in Fig. \ref{fig:kumru4}. 
    Remember that GPEOT-P originally estimates the latent shape by learning the associated projection contours; therefore, to be able to visualize and interpret the results in 3D, we implemented a simple 3D reconstruction algorithm. 
    The algorithm basically starts from a conservative estimate of the underlying 3D volume and refines the estimate by carving out the sections that are inconsistent with the projections. 
    Fig. \ref{fig:kumru4}(b) exhibits the reconstructed shapes as estimates. 
    
    We examine three differently shaped objects (cube, ellipsoid and cone) for the two motion patterns (linear motion and complex maneuver). 
    For each method, all experiments are repeated 100 times with random realizations of the measurement noise and the measurement origins. 
    Table \ref{table:RMSE_vel} reports the averaged RMSE values of the object velocity computed for all simulation experiments. 
    For all experiments, GPEOT and GPEOT-P show superior performance compared to the RM model in terms of velocity estimation. 
    Fig. \ref{fig:kumru5} exhibits the IOU results obtained by averaging the Monte Carlo (MC) runs. 
    GPEOT and GPEOT-P produce successful results for all three shapes while the RM model shows satisfactory performance only for the ellipsoid object. 
    It is an expected finding since both of the GP-based approaches are flexible methods to represent any arbitrary star-convex shape, whereas RM essentially models the underlying shape by an ellipsoid.  
    Additionally, the proposed algorithms are shown to be robust enough to handle the model mismatch in kinematics occurring in the complex maneuvering experiment where the constant velocity model is no longer valid for this motion pattern. 
    A particular reason for their robustness is that they can competently track the orientation of the objects (see Figs. \ref{fig:kumru6} and \ref{fig:kumru7}). 
    Finally, GPEOT is observed to outperform the other algorithms with respect to the IOU measure for all cases. 
    
    \begin{table}[b!]
    	\caption{Root Mean Squared Error (RMSE) of the Object Velocity [$\text{ms}^{-1}$]. (Results are averaged over 100 MC runs.)}
    	\centering
    	\begin{tabular}{|l|c|c|c|c|c|c|}
    		\hline
    		\multicolumn{1}{|c|}{\multirow{2}{*}{}} & \multicolumn{3}{c|}{\textbf{Linear Motion}}        & \multicolumn{3}{c|}{\textbf{Complex Maneuver}}     \\ \cline{2-7} 
    		\multicolumn{1}{|c|}{}                  & \textbf{Cube} & \textbf{Ellipsoid} & \textbf{Cone} & \textbf{Cube} & \textbf{Ellipsoid} & \textbf{Cone} \\ \hline
    		\textbf{GPEOT}                          & 0.124         & 0.125              & 0.150         & 0.112         & 0.120              & 0.124         \\ \hline
    		\textbf{GPEOT-P}                        & 0.162         & 0.184              & 0.200         & 0.203         & 0.184              & 0.192         \\ \hline
    		\textbf{RMM}                            & 0.323         & 0.239              & 0.316         & 0.332         & 0.245              & 0.322         \\ \hline
    	\end{tabular}
    	\label{table:RMSE_vel}
    \end{table}

    \textit{\textbf{Computation time:}} Both of the proposed algorithms are basically realized by an EKF, hence the estimates are recursively updated using newly available measurements at each time step. 
    Therefore, the computational requirements do not increase over time and are basically determined by the size of the state vector and the number of the measurements. 
    The state dimension in GPEOT is $\text{dim}(\mathbf{x}_k) = \text{dim}(\mathbf{c}_k) + \text{dim}(\mathbf{v}_k) + \text{dim}(\mathbf{a}_k) + \text{dim}(\boldsymbol{\omega}_k) + \text{dim}(\mathbf{f}_k) = 654$, and in GPEOT-P, it is $\text{dim}(\mathbf{x}_k) = \text{dim}(\mathbf{c}_k) + \text{dim}(\mathbf{v}_k) + \text{dim}(\mathbf{a}_k) + \text{dim}(\boldsymbol{\omega}_k) + \text{dim}(\mathbf{f}_k^1) + \text{dim}(\mathbf{f}_k^2) + \text{dim}(\mathbf{f}_k^3) = 162$. 
    We utilize a naive implementation of EKF for each method without exploiting any code optimization methods.
    The partial derivatives used in the measurement update phase of the filter are computed numerically. 
    All simulations are conducted in \MAT\ 2017a on a standard laptop with Intel Core i7-6700HQ 2.60 Hz CPU using 16 GB of RAM. 
    Average computation time for an update is recorded as 37.3 ms for GPEOT, 8.2 ms for GPEOT-P and 0.2 ms for RM model. 

    \textit{\textbf{An Alternative Process Model for the Extent:}} 
    To investigate the impact of the extent process model on the performance, we consider an alternative model, which was proposed in \cite{wahlstrom2015extended}: 
    \begin{subequations} \label{eq:extentGPDynamic}
    	\begin{equation}
    	\mathbf{f}_{k+1} = F^\mathbf{f} \mathbf{f}_{k} + \mathbf{w}_k,\quad \mathbf{w}_k \sim \mathcal{N}(\mathbf{0}, Q^\mathbf{f})\ ,
    	\end{equation}
    	where
    	\begin{equation}
    	F^\mathbf{f} = e^{-\alpha T} I,\quad Q^\mathbf{f} = (1-e^{-2\alpha T}) K(\mathbf{u}^\mathbf{f}, \mathbf{u}^\mathbf{f})\ .
    	\end{equation}
    \end{subequations}
    $\alpha$ can be considered as a forgetting factor. 
    Based on this model, we implemented another version of both GPEOT and GPEOT-P for $\alpha=0.0001$. 
    The same simulation experiments are conducted with this implementation repeating 100 MC runs. 
    Tables \ref{table:IOU_linear} and \ref{table:IOU_rotation} present the findings by the mean of the IOU measure for the linear motion and the complex manuevering experiment, respectively. 
    The results suggest that the new implementation also achieves successful performance for all cases, and the alternative process model of the extent does not lead to a significant difference with respect the the IOU measure. 
    
    \begin{table}[b]
    	\caption{Mean of the Intersection-over-Union (IOU) Values for the Linear Motion Experiment. (Results are averaged over 100 MC runs.)}
    	\centering
    	\resizebox{\columnwidth}{!}{
    		\begin{tabular}{|l|l|c|c|c|}
    			\hline
    			\multirow{2}{*}{} & \multirow{2}{*}{\textbf{Process Model for the Extent}} & \multicolumn{3}{c|}{\textbf{Object Shape}} \\ \cline{3-5} 
    			&  & \textbf{Cube} & \textbf{Ellipsoid} & \textbf{Cone} \\ \hline
    			\multirow{2}{*}{\textbf{GPEOT}} & Maximum Entropy Model & 0.908 & 0.910 & 0.824 \\ \cline{2-5} 
    			& Forgetting Factor Model & 0.906 & 0.914 & 0.832 \\ \hline
    			\multirow{2}{*}{\textbf{GPEOT-P}} & Maximum Entropy Model & 0.818 & 0.829 & 0.741 \\ \cline{2-5} 
    			& Forgetting Factor Model & 0.816 & 0.826 & 0.737 \\ \hline
    	\end{tabular}}
    	\label{table:IOU_linear}
    \end{table}
    
    \begin{table}[b]
    	\caption{Mean of the Intersection-over-Union (IOU) Values for the Complex Maneuvering Experiment. (Results are averaged over 100 MC runs.)}
    	\centering
    	\resizebox{\columnwidth}{!}{
    		\begin{tabular}{|l|l|c|c|c|}
    			\hline
    			\multirow{2}{*}{} & \multirow{2}{*}{\textbf{Process Model for the Extent}} & \multicolumn{3}{c|}{\textbf{Object Shape}} \\ \cline{3-5} 
    			&  & \textbf{Cube} & \textbf{Ellipsoid} & \textbf{Cone} \\ \hline
    			\multirow{2}{*}{\textbf{GPEOT}} & Maximum Entropy Model & 0.897 & 0.907 & 0.866 \\ \cline{2-5} 
    			& Forgetting Factor Model & 0.896 & 0.909 & 0.865 \\ \hline
    			\multirow{2}{*}{\textbf{GPEOT-P}} & Maximum Entropy Model & 0.782 & 0.788 & 0.718 \\ \cline{2-5} 
    			& Forgetting Factor Model & 0.786 & 0.786 & 0.715 \\ \hline
    	\end{tabular}}
    	\label{table:IOU_rotation}
    \end{table}
    
    \textit{\textbf{Effect of Center Initialization:}} 
    As mentioned earlier, the proposed extent models do not rely on a unique definition of the center point. 
    Instead, different points within the object extent can be specified to be the center position, and together with compatible radial functions, these can describe precisely the same object. 
    Therefore, a possible area for future research would be to develop mechanisms to initialize the center point that maximize the performance of the suggested trackers. 
    
\begin{figure}[t]
    	\centering
    	\includegraphics[trim= 30 0 105 15,clip, width=0.95\columnwidth]{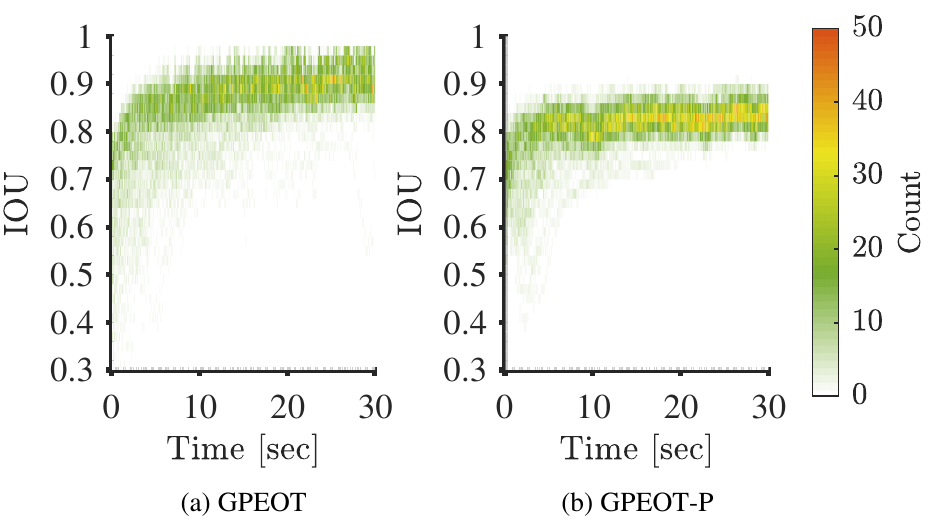}
    	\caption{Histogram plots of the Intersection-over-Union (IOU) measure calculated for 100 MC runs of the linear motion experiment with randomly initialized center points. }
    	\centering
    	\label{fig:kumru8}
    \end{figure}

    In this regard, we examine the sensitivity of the proposed algorithms to the initialization of the center point.
    We consider the linear motion experiment for the cube-shaped object. 
    The simulation is repeated for 100 times, and at each run, the initial center point is sampled from a uniform distribution defined within the entire volume of the object. 
    The results are illustrated by the histograms that visualize the complete distribution of the IOU values at each time instant of the scenario in Fig. \ref{fig:kumru8}. 
    Both of the algorithms are shown to be robust against center initialization, as they consistently converge to high IOU values with decreasing variance throughout the experiment.
    
\begin{figure}[t]
    	\centering
    	\includegraphics[trim= 10 5 15 0,clip, width=0.85\columnwidth]{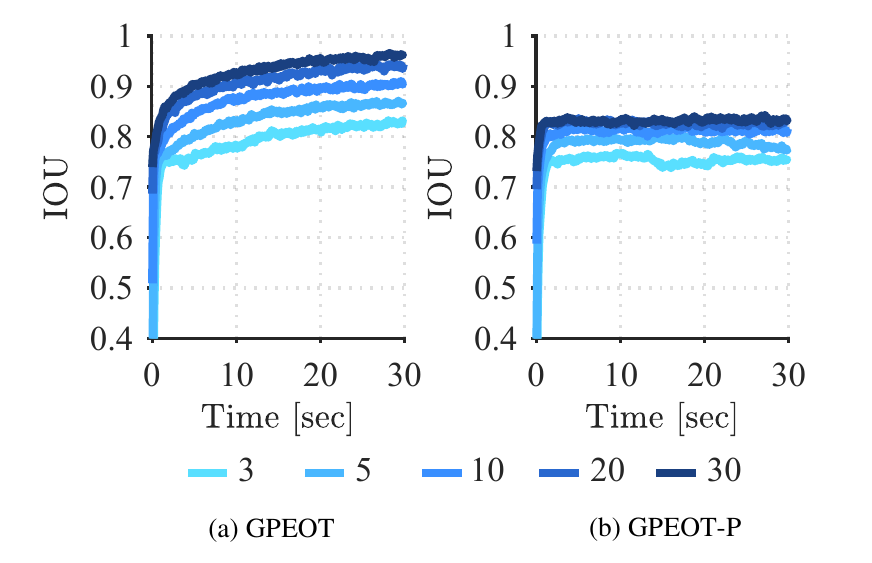}
    	\caption{Intersection-over-Union (IOU) plots obtained for different number of available measurements per frame. (The results are averaged over 100 MC runs.)}
    	\centering
    	\label{fig:kumru9}
    \end{figure}

    \textit{\textbf{Effect of Number of Measurements per Frame:}} 
    In this subsection, we investigate how the performance of the proposed methods depends on the number of available measurements per frame. 
    To this end, we consider the linear motion experiment for the cube-shaped object. 
    We repeat the experiment for the following number of measurements per frame: ${\{3, 5, 10, 20, 30\}}$. 
    The simulations are run 100 times for each case, where the measurement sources are randomly sampled from the surface and corrupted by independent measurement noise. 
    Fig. \ref{fig:kumru9} reports the averaged IOU measure for all experiments. 
    Both methods perform satisfactorily even when the number of measurements is relatively low.  
    Additionally, the performance of the methods regularly improves with the increasing number of measurements,
    which indicates that they can effectively assimilate information provided by the measurements. 
    
    
    \begin{figure}[b]
    	\centering
    	\includegraphics[trim= 0 0 0 0,clip, width=0.9\columnwidth]{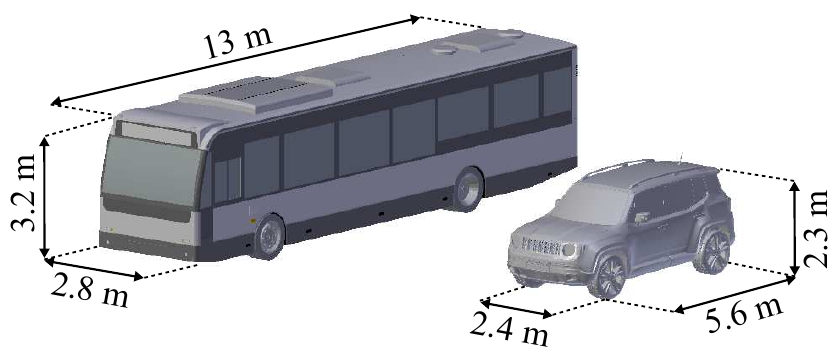}
    	\caption{Realistic vehicle models utilized in the Blensor experiments. }
    	\centering
    	\label{fig:kumru10}
    \end{figure}
    \begin{figure}[t]
    	\centering
    	\subfloat[GPEOT]{\includegraphics[trim= 0 0 0 0,clip, width=0.49\columnwidth]{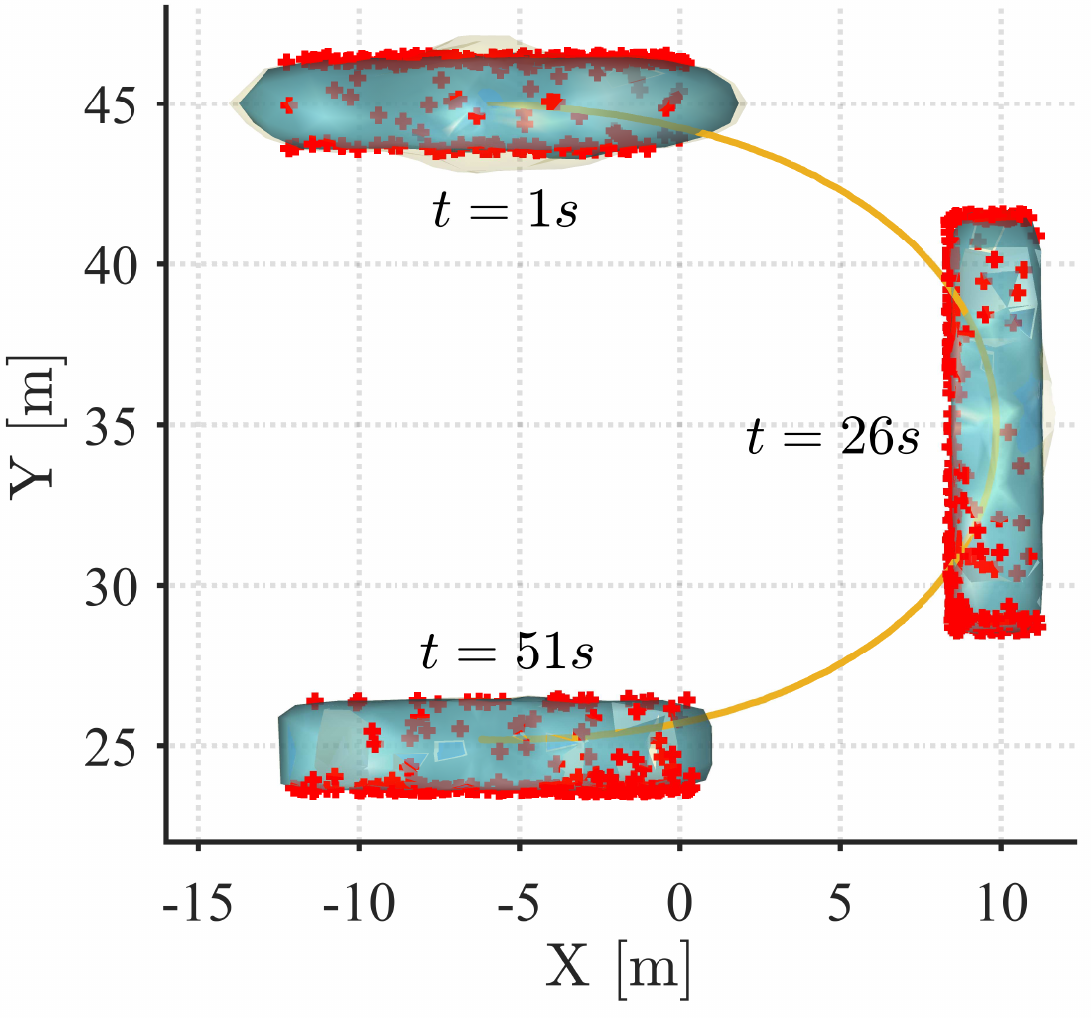}}
    	\subfloat[GPEOT-P]{\includegraphics[trim= 0 0 0 0,clip, width=0.49\columnwidth]{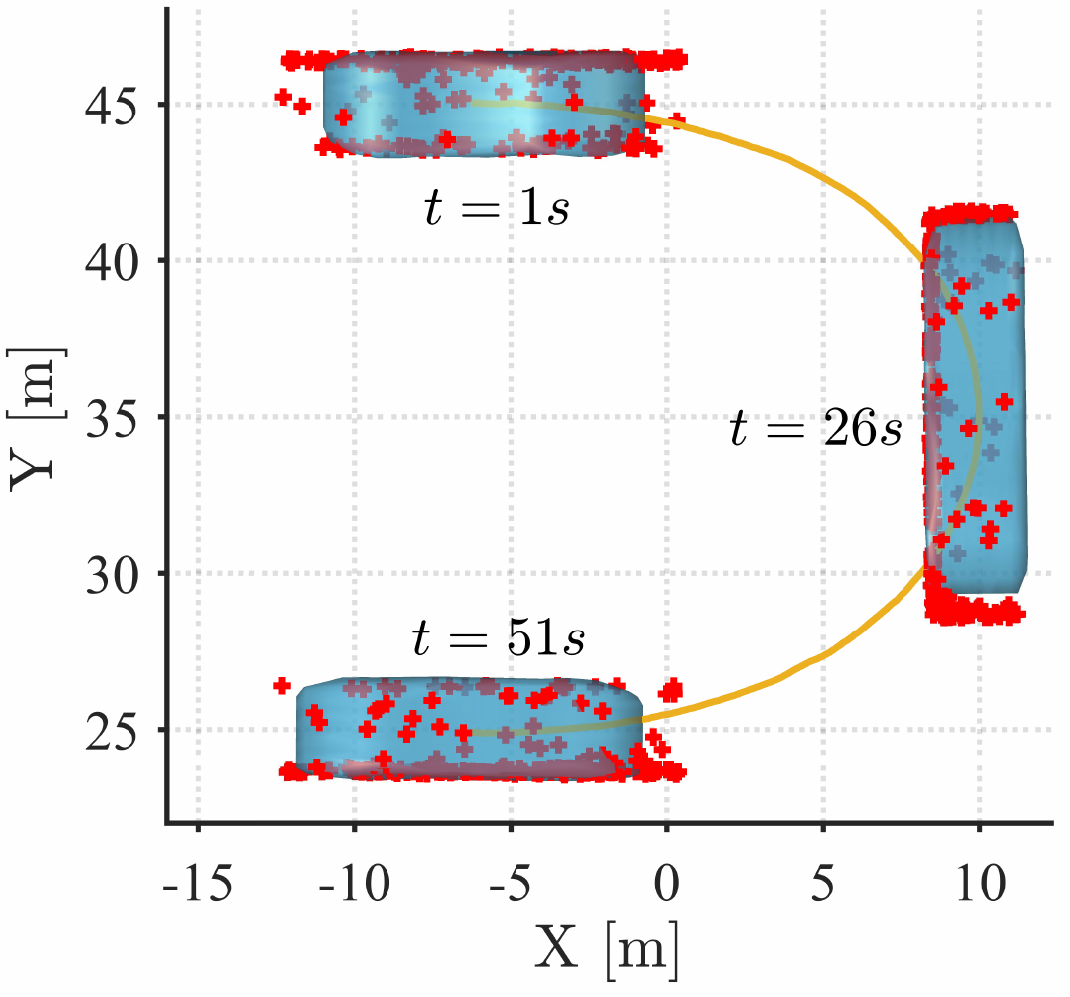}}\\
    	\subfloat[GPEOT]{\includegraphics[trim= 0 0 0 0,clip, width=0.49\columnwidth]{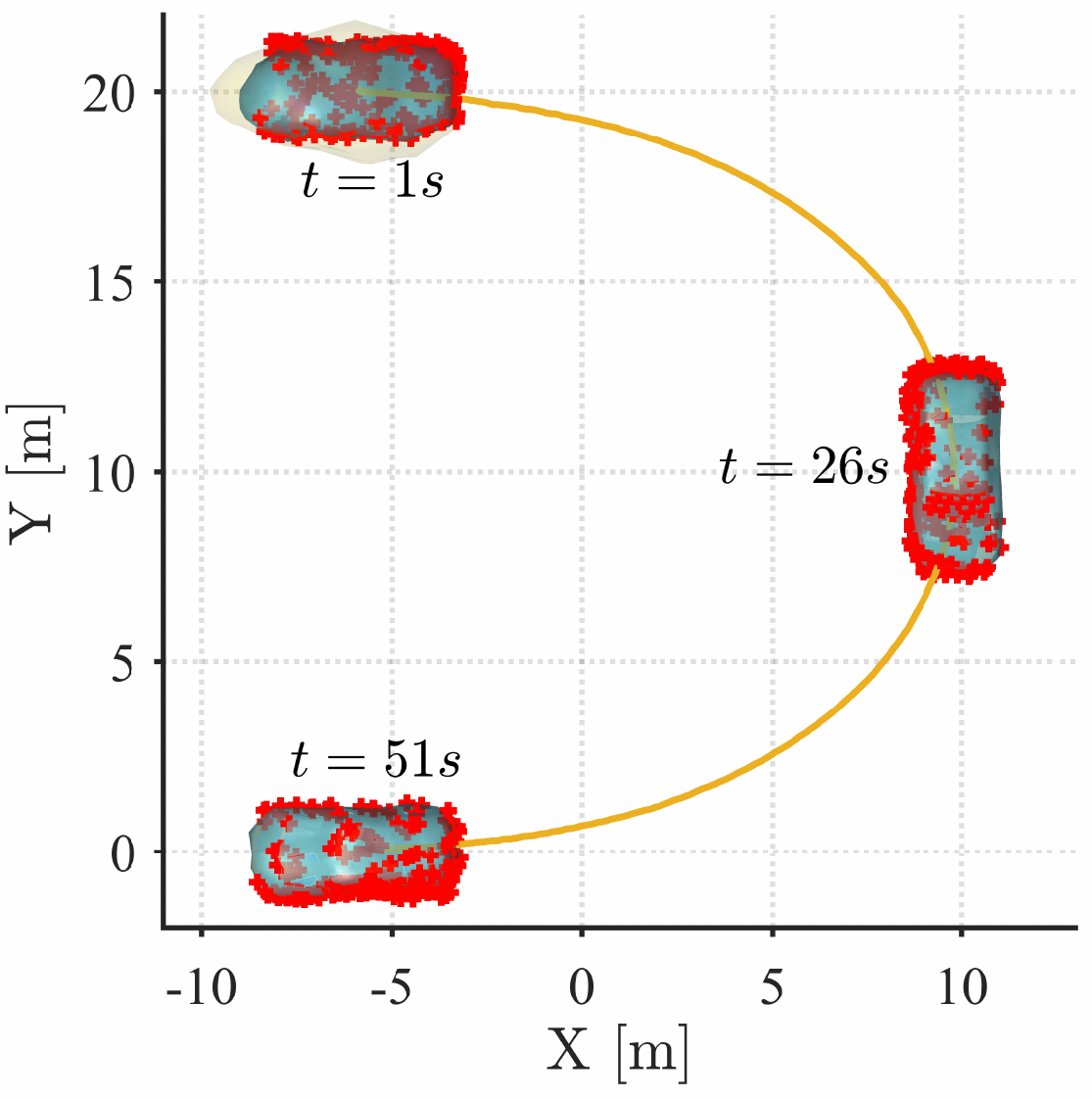}}
    	\subfloat[GPEOT-P]{\includegraphics[trim= 0 0 0 0,clip, width=0.49\columnwidth]{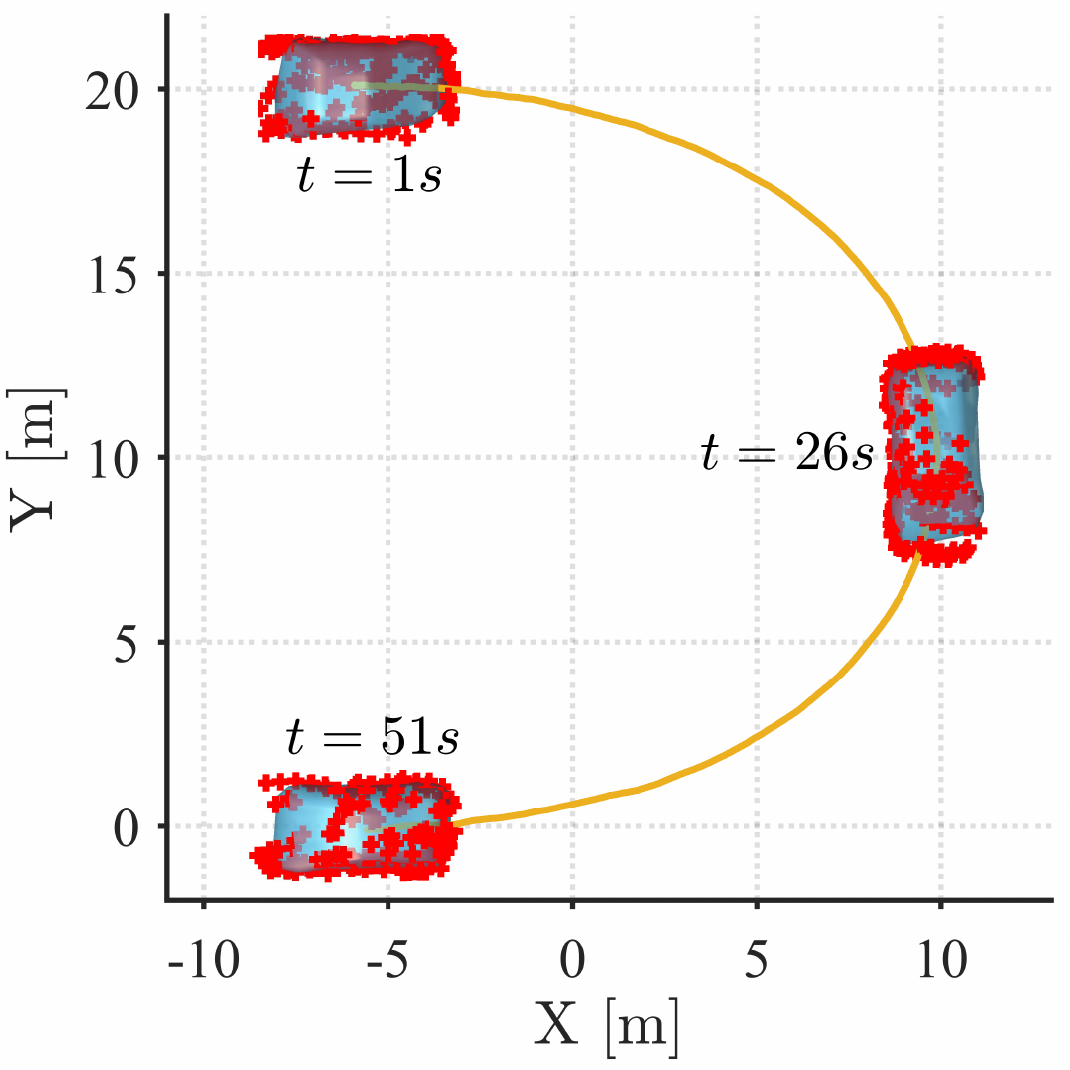}}
    	\caption{Results obtained during Blensor simulations. 
    		(In Figs. (a) and (b), the bus is observed by two sensors at (0, 60, -5) and (0, 15, 5). 
    		In Figs. (c) and (d), the jeep is observed by two sensors at (0, 30, -5) and (0, -10, 5).)}
    	\label{fig:kumru11}
    \end{figure}
    
    \subsubsection{Blensor Simulations} \label{sec:Result_Blensor}
    To qualify the representational power of the suggested algorithms, additional experiments are conducted in Blensor, which is a high fidelity sensor simulation environment. 
    In these experiments, we consider realistic models of two different types of vehicles, namely a bus and a jeep, which are depicted in Fig. \ref{fig:kumru10}. 
    In the scenario, each vehicle makes a u-turn while being observed by two Velodyne HDL-64E2 LIDAR sensors. 
    The parameters of the algorithms are kept the same as in the previous subsection with two exceptions: the covariance matrix of rotational process noise is set to ${\Sigma_\alpha = \text{diag}(0,0,\sigma_\alpha^2)}$ to reflect that the considered vehicles can only rotate around their yaw-axis, and the length-scale of GPEOT is set to $l = \pi/12$. 
    \begin{figure}[t]
    	\centering
    	\subfloat[GPEOT]{\includegraphics{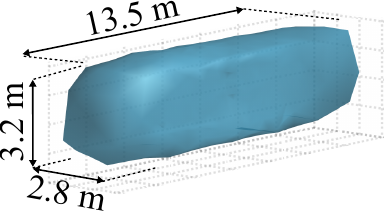} 
    		\includegraphics{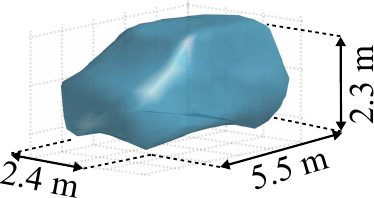}}\\
    	\subfloat[GPEOT-P]{\includegraphics{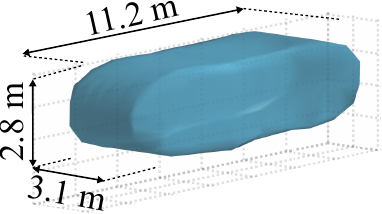}
    		\includegraphics{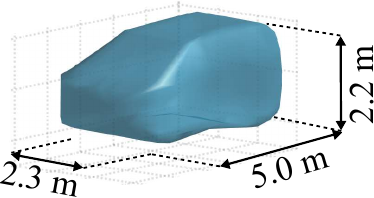}}
    	\caption{Close-up views of the extent estimates obtained at the last instant of the Blensor simulations. }
    	\label{fig:kumru12}
    \end{figure}
    \begin{figure}[t]
    	\centering
    	\subfloat[GPEOT]{\includegraphics[trim= 0 0 0 0,clip]{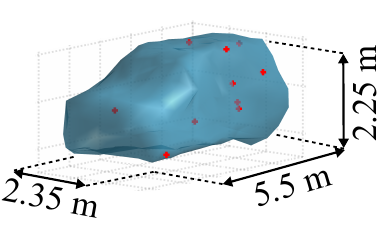}}\hspace{5pt}
    	\subfloat[GPEOT-P]{\includegraphics[trim= 0 0 0 0,clip]{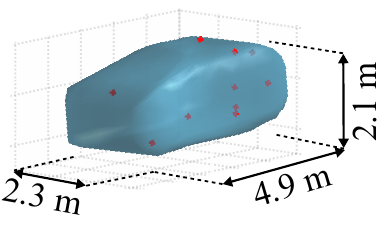}}
    	\caption{Close-up views of extent estimates obtained at the last instant of the Blensor experiment, in which only 10 measurements per frame are provided to the algorithms.}
    	\centering
    	\label{fig:kumru13}
    \end{figure}

    The overview of the tracking outputs is shown in Fig. \ref{fig:kumru11}. 
    Both of the proposed methods can successfully track the two different vehicles. 
    Furthermore, the shape estimates obtained at the last instant of the experiments are demonstrated by some close-up views in Fig. \ref{fig:kumru12}. 
    While GPEOT is able to capture a highly detailed representation of the underlying object extent, GPEOT-P achieves a satisfactory but rather rough shape estimate. 
    Besides, GPEOT-P slightly underestimates the size of the object due to the mismatch between the specified and true values of the scaling factor used in the measurement model. 
    
    Finally, to further investigate the effect of the number of measurements on the estimation performance, we consider the same Blensor experiment with the jeep vehicle. 
    The employed LIDAR sensors are able collect an abundant number of measurements returned from the object of interest, which is well over a thousand for the most sensor scans. 
    For our purposes, we randomly select 10 measurements from the acquired point cloud at each frame and provide them to the algorithms. 
    The resulting extent estimates obtained at the end of the scenario are illustrated in Fig. \ref{fig:kumru13}. 
    Both of the algorithms estimate the kinematics and the extent satisfactorily during the experiment. 
    In particular, GPEOT can successfully form such a detailed description of the underlying shape with only 10 measurements per frame. 
    In this regard, the proposed method does not require the substantial amount of information generated by the LIDAR sensors; instead, it can effectively perform with significantly fewer measurements. 
    
    \subsection{Experiments with Real Data} \label{sec:Result_RealData}
    In this section, the performance of the algorithms is assessed on real data. 
    To this end, we hereby use the Kitti tracking benchmark, \cite{Geiger2012CVPR}. 
    The benchmark consists of various records of real-world traffic scenarios captured by several sensor modalities mounted on an ego vehicle. 
    We form two scenarios of different vehicles by extracting the corresponding sequences of point measurements acquired by a Velodyne HDL-64E laser scanner. 
    Note that we do not consider the preprocessing of the raw point cloud data, e.g., ground removal, segmentation, association, as it is beyond the scope of this work; instead, the sequences are extracted using the labels provided in the benchmark. 
    The same sets of parameters as in the previous subsections are utilized except the length-scale of GPEOT is set to $l = \pi/14$. 
    
    \begin{figure}[t]
    	\centering
    	\subfloat[Scenario 1]{\includegraphics[trim= 0 0 0 0,clip, width=0.495\columnwidth]{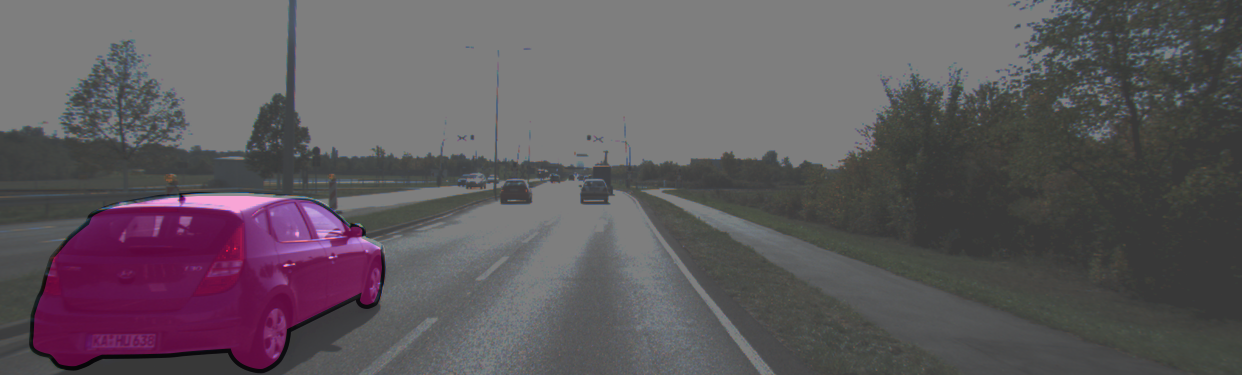}\hspace{-2pt}
    		\includegraphics[trim= 0 0 0 0,clip, width=0.495\columnwidth]{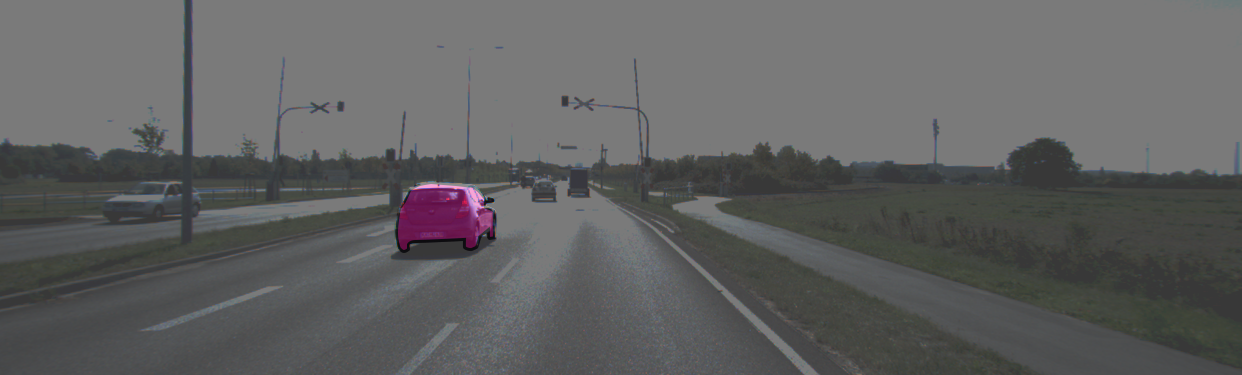}}\\
    	\subfloat[Scenario 2]{\includegraphics[trim= 0 0 0 0,clip, width=0.495\columnwidth]{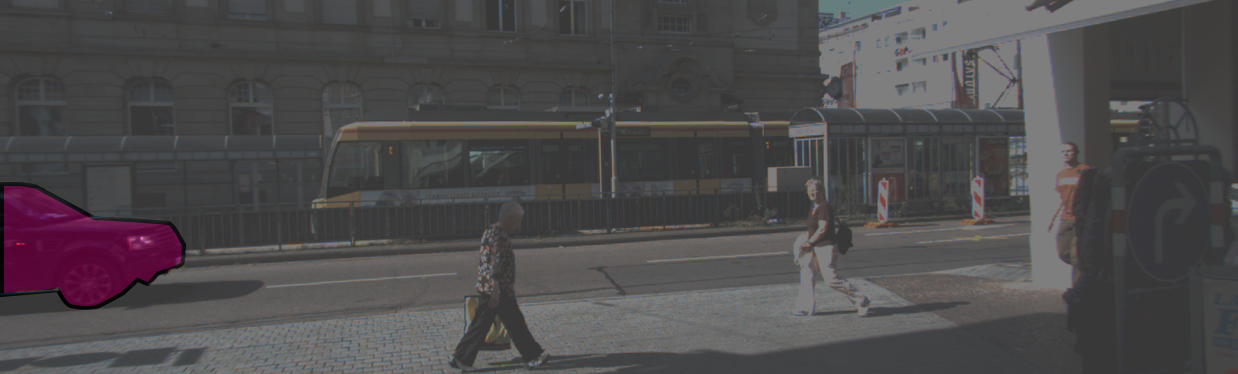}\hspace{-2pt}
    		\includegraphics[trim= 0 0 0 0,clip, width=0.495\columnwidth]{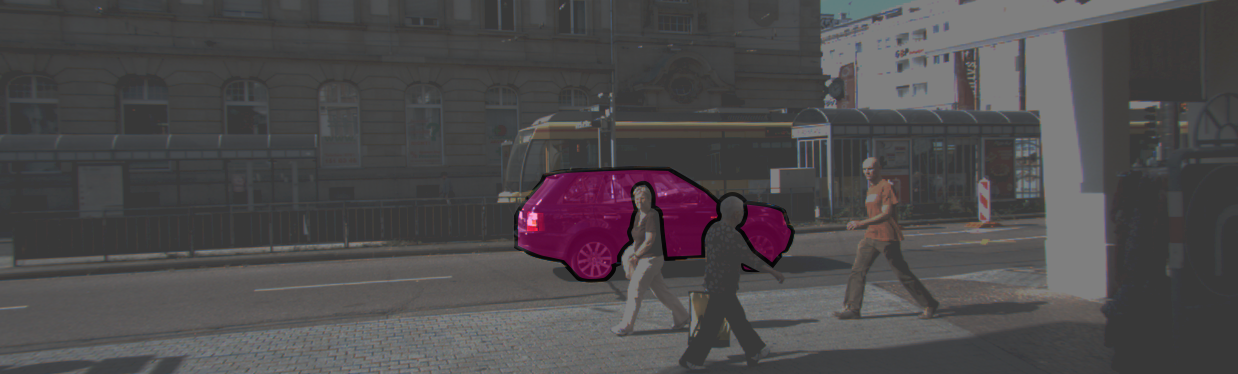}}
    	\caption{Example views captured by a camera mounted next to the laser scanner on the ego vehicle. Left and right images depict the initial and the intermediate frames of the scenarios, respectively. Note that the highlighted vehicles are tracked by the proposed algorithms using only point cloud measurements.}
    	\label{fig:kumru14}
    \end{figure}
    
    \begin{table}[t!]
    	\caption{Root Mean Squared Error (RMSE) of the Center [m] and the Yaw Angle [degree] in the Real Data Experiments.}
    	\centering
        \begin{tabular}{|l|c|c|c|c|}
        \hline
        \multirow{2}{*}{} & \multicolumn{2}{c|}{\textbf{Scenario 1}} & \multicolumn{2}{c|}{\textbf{Scenario 2}} \\ \cline{2-5} 
                          & \textbf{Center}   & \textbf{Yaw Angle}  & \textbf{Center}   & \textbf{Yaw Angle}      \\ \hline
        \textbf{GPEOT}    & 0.21                  & 1.37                 & 0.23                      & 1.40                     \\ \hline
        \textbf{GPEOT-P}  & 0.27                  & 3.81                 & 0.25                      & 1.90                     \\ \hline
        \end{tabular}
    	\label{table:Kitti_yawRMSE2}
    \end{table}
    
    \begin{figure} [t]
    	\centering
    	\subfloat[GPEOT]{\includegraphics[trim= 0 0 0 0,clip, width=\columnwidth]{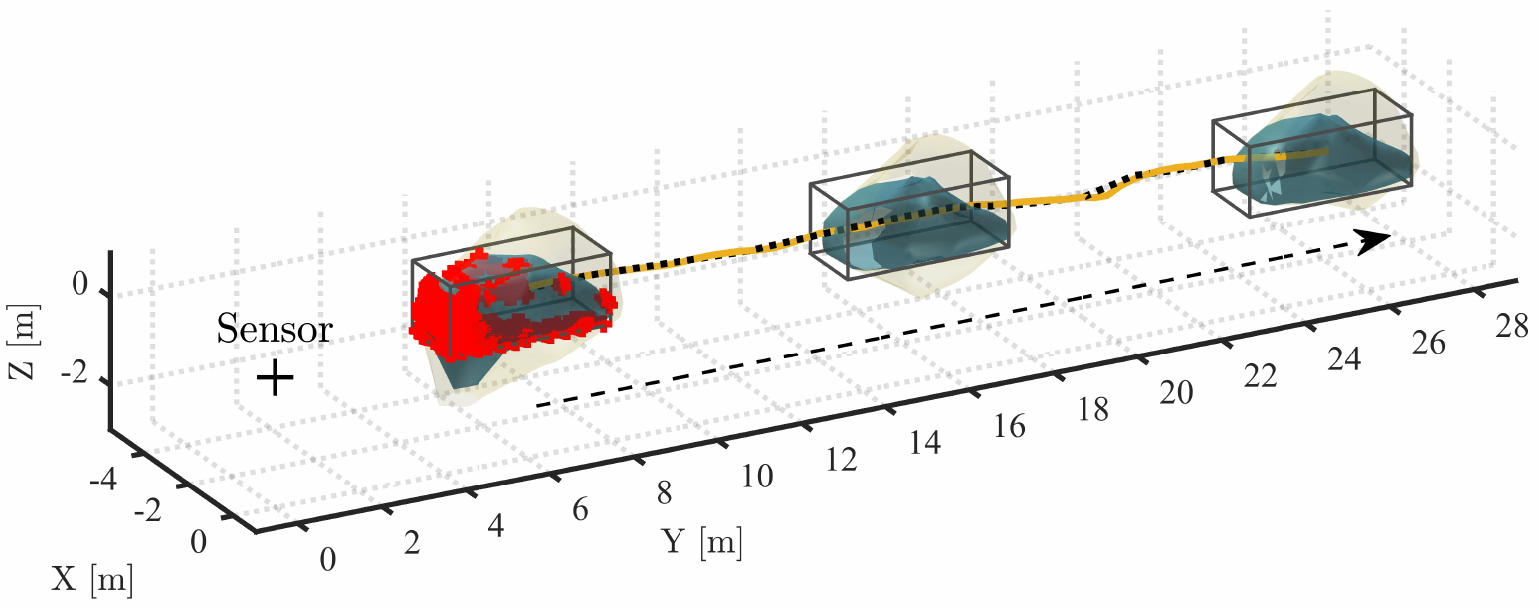}}\\
    	\subfloat[GPEOT-P]{\includegraphics[trim= 0 0 0 0,clip, width=\columnwidth]{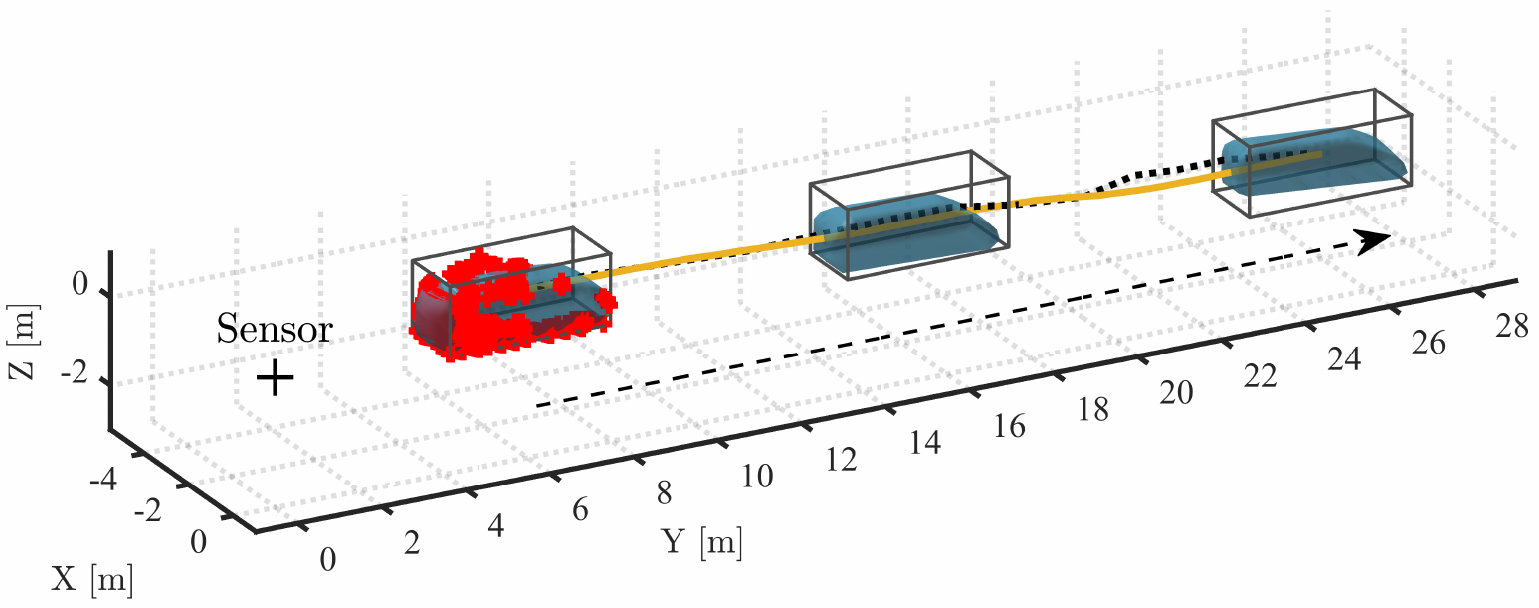}}
    	\caption{Scenario 1 with real data. 
    		(Blue and yellow surfaces indicate the estimated extent and the predicted uncertainty of one standard deviation, respectively. 
    		Bounding box denotes the ground truth annotation of the target. 
    		Red plus signs plotted for the first frame visualize the measurements. 
    		Solid yellow and dashed black curves are the estimated and true trajectory, respectively. 
    		Dashed black arrow is the direction of the target.)}
    	\label{fig:kumru15}
    \end{figure}
    \begin{figure} [t]
    	\centering
    	\subfloat[GPEOT]{\includegraphics[trim= 0 0 0 0,clip, width=0.8\columnwidth]{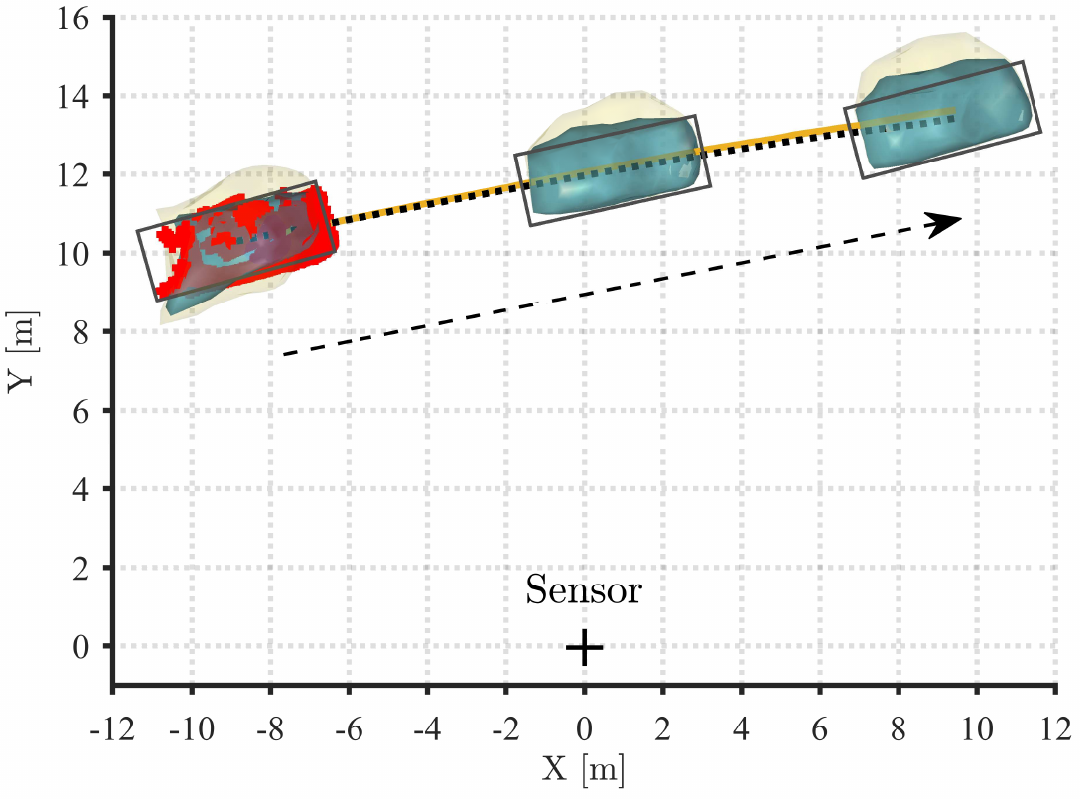}}\\
    	\subfloat[GPEOT-P]{\includegraphics[trim= 0 0 0 0,clip, width=0.8\columnwidth]{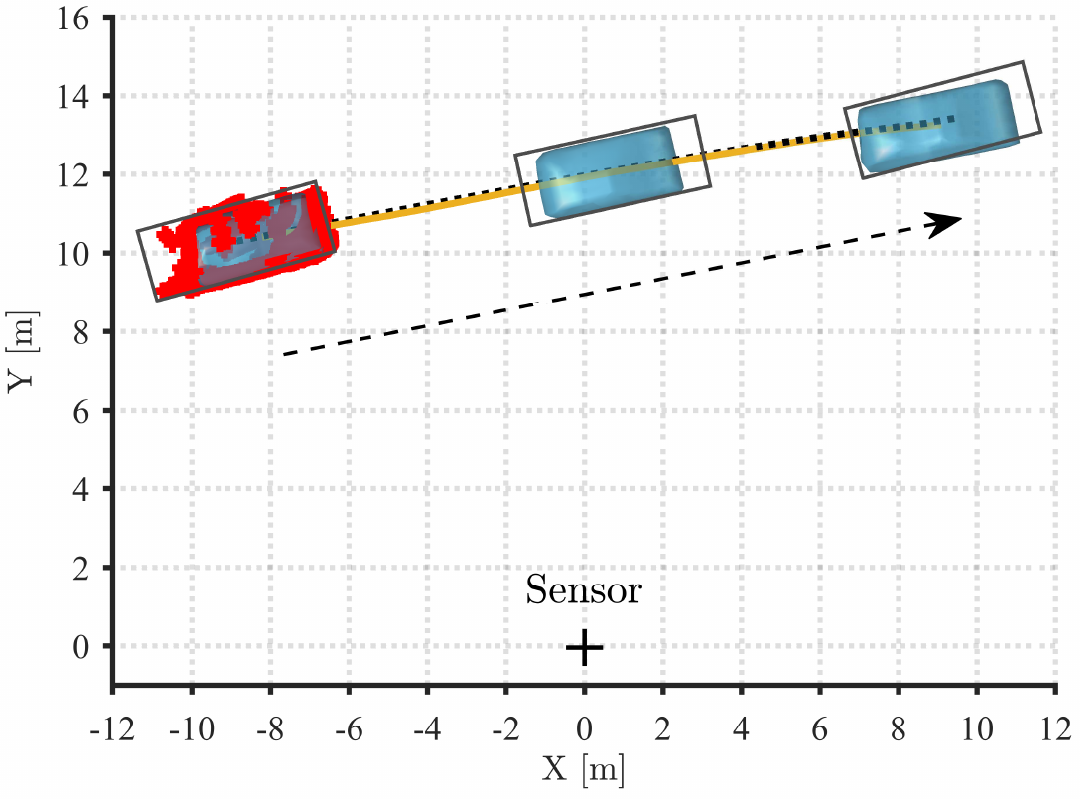}}
    	\caption{Scenario 2 with real data. 
    	(Blue and yellow surfaces indicate the estimated extent and the predicted uncertainty of one standard deviation, respectively. 
    	Bounding box denotes the ground truth annotation of the target. 
    	Red plus signs plotted for the first frame visualize the measurements. 
    	Solid yellow and dashed black curves are the estimated and true trajectory, respectively. 
    	Dashed black arrow is the direction of the target.)}
    	\label{fig:kumru16}
    \end{figure}

    The scenarios are visualized in Fig. \ref{fig:kumru14}. 
    The first scenario takes place on a highway where the ego and the target vehicle move in the same direction, and the target pulls consistently ahead in time. 
    The dataset provides the ground truth in terms of bounding boxes; hence, to assess the algorithms' performance quantitatively, we fit bounding boxes to the estimated extents and compute the RMSE of the center and the yaw angle accordingly. 
    Table \ref{table:Kitti_yawRMSE2} presents the corresponding results. 
    Fig. \ref{fig:kumru15} demonstrates that both of the algorithms accomplish successful tracking. 
    Throughout the experiment, the sensor can only observe the back and right side of the target, thus the uncertainty of the extent on the observed portion decreases in time while a high uncertainty is properly associated with the unobserved section as explicitly shown in Fig. \ref{fig:kumru15}. 
    The GPEOT-P implementation uses the periodic covariance function in \eqref{eq:covFuncSym} for the projection onto the ground plane. 
    Considering solid amount of empirical evidence, this is a reasonable assumption for many targets in driving settings. 
    The implementation inherently assumes that the corresponding projection contour is periodic with $\pi$ so that the radial function takes exactly same values for $f(\theta)$ and $f(\theta+\pi)$. 
    Accordingly, the reconstructed shape estimates accurately captures the appearance on the unobserved section of the object as seen in Fig. \ref{fig:kumru15}. 
    
    In the second scenario, the ego vehicle waits stationary at a road junction while the target vehicle crosses the street.
    The experiment imposes two main challenges: 
    First, the target is temporarily occluded by pedestrians and a column of a building; second, there are respectable number of 3D point measurements returned from the driver and the interior structure of the target vehicle.  
    The tracking outputs of the algorithms and the corresponding RMSE values are presented in Fig. \ref{fig:kumru16} and Table \ref{table:Kitti_yawRMSE2}, respectively. 
    GPEOT-P makes use of the periodic covariance function for the ground projection as in the previous case. 
    Both of the methods achieve accurate tracking and prove their robustness against occlusions and interior measurements. 
	
    \section{Conclusion and Future Work} \label{sec:Conclusion}
    A new approach is proposed to contribute to the perception of autonomous systems. 
    The proposed algorithm is capable of processing 3D point cloud data for tracking dynamic objects with unknown shapes. The method can exploit the full potential of the information hidden in, possibly sparse, point cloud measurements by estimating the object's shape simultaneously with its kinematic state including the position, velocity and orientation. 
    The proposed model is flexible to express and learn a large variety of shapes which may co-exist in a surveillance region. 
    An alternative efficient implementation of the method is also derived, which reduces the computational requirements by utilizing plane projections. 
    The algorithms are efficient in the implementation such that an extension to the multi-target tracking framework and real-time applications are possible. 
    The methods provide an analytical expression of the object's shape, and this information can later be used for online identification and classification purposes in the future. 
    The performance of the suggested approach can further be improved by using negative information embedded in the point measurements, considering alternative covariance functions and inference methods, and incorporating dedicated models for the scaling factor in the projections. 
    The methods are applicable to various fields which require accuracy in perception such as the robotics applications where a moving agent is required to navigate in a 3D environment with 3D motion constraints or offline applications such as 3D map extraction.

	
	\appendices
	\section{Recursive Gaussian Process Regression for Arbitrary Mean Functions} \label{sec:RGP_ArbitraryMean}
	In this section, we will give the equations of recursive GP regression for models with arbitrary mean functions, i.e., 
	${f(u) \sim \mathcal{GP}(\mu(u), k(u,u'))}$ where there is no restriction on $\mu(\cdot)$. 
	Based on this model, the joint distribution of a measurement, $m_k$, and the latent function values, $\mathbf{f}$, is described by 
	\begin{equation}
	\begin{bmatrix}  \nonumber
	m_k\\ 
	\mathbf{f}
	\end{bmatrix} \sim \mathcal{N}
	\left(
	\begin{bmatrix}
	\mu(u_k)\\
	\boldsymbol{\mu}(\mathbf{u}^\mathbf{f})
	\end{bmatrix},
	\begin{bmatrix}
	k(u_k, u_k)+R &  K(u_k, \mathbf{u}^\mathbf{f})\\
	K(\mathbf{u}^\mathbf{f}, u_k) & K(\mathbf{u}^\mathbf{f}, \mathbf{u}^\mathbf{f})
	\end{bmatrix}
	\right).
	\end{equation}
	
	Then, the conditional density of $m_k$ becomes, \cite[Ch. 2.7]{rasmussen2006gaussian}
	\begin{subequations} \label{eq:GP_ArbitraryMean}  
		\begin{align}
		p(m_k|\mathbf{f}) &= \mathcal{N}(m_k; H_k^\mathbf{f}\mathbf{f} + c(u_k), R_k^\mathbf{f}),
		\end{align}
		where
		\begin{align} \label{eq:cDef}
		c(u_k) =\mu(u_k) - H^\mathbf{f}_k \boldsymbol{\mu}(\mathbf{u}^\mathbf{f}),
		\end{align}
		and the definitions of $H^\mathbf{f}_k$ and $R^\mathbf{f}_k$ remain the same as given in \eqref{eq:recursiveGPFormula}
		\begin{align}
		H^\mathbf{f}_k &= H^\mathbf{f}(u_k) =  K(u_k, \mathbf{u}^\mathbf{f}) [K(\mathbf{u}^\mathbf{f}, \mathbf{u}^\mathbf{f})]^{-1}, \nonumber \\
		R^\mathbf{f}_k &= R^\mathbf{f}(u_k) = k(u_k, u_k)+R \nonumber\\
		& \quad - K(u_k, \mathbf{u}^\mathbf{f}) [K(\mathbf{u}^\mathbf{f}, \mathbf{u}^\mathbf{f})]^{-1} K(\mathbf{u}^\mathbf{f}, u_k).\nonumber
		\end{align}
	\end{subequations}
	
	Similarly to the zero-mean case, \eqref{eq:GP_ArbitraryMean} allows us to construct a state space model, which may be regarded by a Kalman filter for recursive inference. 
	The sole difference of the resulting model from the one given in \eqref{eq:recursiveGPFormula} is that the measurement model now includes an additional term $c(u_k)$ defined in \eqref{eq:cDef} as  
	\begin{align}
	m_k &= H^\mathbf{f}_k\ \mathbf{f}_{k} + c(u_k) + e_k^\mathbf{f},\quad e_k^\mathbf{f} \sim \mathcal{N}(0, R^\mathbf{f}_k), 
	\end{align}
	where $\mathbf{f}_{k} = \mathbf{f}$.

	\section{Extended Kalman Filtering Using Pseudo Measurements} \label{sec:App_EKF}
	Both of the algorithms proposed in the paper, perform inference by an EKF. 
	The filter recursively updates the estimate and the associated covariance after they are initialized as $\hat{\mathbf{x}}_{0|-1} = \boldsymbol{\mu}_0$,  $P_{0|-1} = P_0$.
	\subsubsection{Measurement Update} \label{sec:App_EKF_meas}
	The equations for the measurement update are as follows. 
	\begin{subequations}
		\begin{align}	
		\hat{\mathbf{x}}_{k|k} &= \hat{\mathbf{x}}_{k|k-1} + K_k (\mathbf{0} - \hat{\mathbf{m}}_{k|k-1})\label{eq:EKFMeanUpdate} \\
		P_{k|k} &= P_{k|k-1} - K_k H_k P_{k|k-1}
		\end{align}
		where
		\begin{align}
		\hat{\mathbf{m}}_{k|k-1} &= \mathbf{h}_k(\mathbf{m}_{k}, \hat{\mathbf{x}}_{k|k-1})\\
		K_k &= P_{k|k-1} H^\top_k S_k^{-1} \\
		S_k &= H_k P_{k|k-1} H^\top_k  + R_k \\ 
		H_k &= \frac{d}{d\mathbf{x}_k} \mathbf{h}_k(\mathbf{m}_{k}, \mathbf{x}_k) |_{\mathbf{x}_k = \hat{\mathbf{x}}_{k|k-1}}
		\end{align}
	\end{subequations}
	Notice that this implementation slightly differs from a standard EKF as the pseudo-measurements, indicated by the zero vector in \eqref{eq:EKFMeanUpdate}, are used in accordance with the derived measurement models. 
	
	\subsubsection{Time Update}
	The time update is performed in a standard manner as in 
	\begin{subequations}
		\begin{align}
		\hat{\mathbf{x}}_{k+1|k} &= F_k \hat{\mathbf{x}}_{k|k}, \\
		P_{k+1|k} &= F_k P_{k|k} F_k^\top + Q_k.
		\end{align}
	\end{subequations}
	
	\section{Details of the Matrices Used in the Rotational Motion Model} \label{sec:App_RotDyn}
	The state space model for describing the dynamics of the rotational motion is given in \eqref{eq:DTRotDyn}. 
	In this section, we reveal the details of the matrices used in the model. 
	\begin{subequations}
		\begin{align}
		F^r_k &= \exp(A_k^r T) \nonumber \\ 
		&= \begin{bmatrix}
		\exp (\frac{T}{2}[-\hat{\boldsymbol{\omega}}_{k|k}\times]) & T \exp(\frac{T}{2}[-\hat{\boldsymbol{\omega}}_{k|k}\times]) \\ 
		0_3 & I_3
		\end{bmatrix}
		\end{align}
		where 
		\begin{align}
		\exp(\frac{T}{2}[-\hat{\boldsymbol{\omega}}_{k|k}\times]) &= I_3 
		+ \frac{\sin(\frac{T}{2} |\hat{\boldsymbol{\omega}}_{k|k}| )}{|\hat{\boldsymbol{\omega}}_{k|k}|} [-\hat{\boldsymbol{\omega}}_{k|k}\times]
		\nonumber \\ &+ \frac{1-\cos(\frac{T}{2} |\hat{\boldsymbol{\omega}}_{k|k}| )}{|\hat{\boldsymbol{\omega}}_{k|k}|^2}  [-\hat{\boldsymbol{\omega}}_{k|k}\times]^2
		\end{align}
	\end{subequations}
	$|\cdot|$ indicates the Euclidean norm. 
	
	\begin{subequations}
		\begin{align}
		G_k &= \left(\int_{0}^{T} \exp(A_k^r \tau) d\tau \right) B \nonumber \\
		&= \begin{bmatrix}
		\int_{0}^{T} \exp (\frac{\tau}{2}[-\hat{\boldsymbol{\omega}}_{k|k}\times]) d\tau & 
		\int_{0}^{T} \tau \exp (\frac{\tau}{2}[-\hat{\boldsymbol{\omega}}_{k|k}\times]) d\tau \\
		0_3  &  \int_{0}^{T} I_3 d\tau
		\end{bmatrix} B
		\end{align}
		where
		\begin{align} 
		\int_{0}^{T} \exp (\frac{\tau}{2}[-\hat{\boldsymbol{\omega}}&_{k|k}\times]) d\tau  = 
		T I_3 \nonumber \\
		&+ \frac{2 (1-\cos(\frac{T}{2} |\hat{\boldsymbol{\omega}}_{k|k}| ))}{|\hat{\boldsymbol{\omega}}_{k|k}|^2} [-\hat{\boldsymbol{\omega}}_{k|k}\times] \nonumber \\
		& + \frac{T-\frac{2}{|\hat{\boldsymbol{\omega}}_{k|k}|}\sin(\frac{T}{2} |\hat{\boldsymbol{\omega}}_{k|k}| )}{|\hat{\boldsymbol{\omega}}_{k|k}|^2} [-\hat{\boldsymbol{\omega}}_{k|k}\times]^2  
		\end{align}
		\begin{align}
		&\int_{0}^{T} \tau \exp (\frac{\tau}{2}[-\hat{\boldsymbol{\omega}}_{k|k}\times]) d\tau = 
		\frac{T^2}{2} I_3 \nonumber \\ 
		& + 
		\frac{1}{|\hat{\boldsymbol{\omega}}_{k|k}|^2} \left(\frac{4}{|\hat{\boldsymbol{\omega}}_{k|k}|} \sin(\frac{T}{2} |\hat{\boldsymbol{\omega}}_{k|k}| ) - 2 T \cos(\frac{T}{2} |\hat{\boldsymbol{\omega}}_{k|k}| )\right) [-\hat{\boldsymbol{\omega}}_{k|k}\times]\nonumber \\	 
		& + \frac{1}{|\hat{\boldsymbol{\omega}}_{k|k}|^2} \left(  \frac{T^2}{2} + \frac{2T}{|\hat{\boldsymbol{\omega}}_{k|k}|} \sin(\frac{T}{2} |\hat{\boldsymbol{\omega}}_{k|k}| ) \right.\nonumber \\
		&\qquad\qquad + \left. \frac{4}{|\hat{\boldsymbol{\omega}}_{k|k}|^2} (\cos(\frac{T}{2} |\hat{\boldsymbol{\omega}}_{k|k}| ) - 1 )\right) [-\hat{\boldsymbol{\omega}}_{k|k}\times]^2
		\end{align}
		\begin{align}
		\int_{0}^{T} I_3 d\tau = T I_3
		\end{align}
	\end{subequations}

	\section{Update of the Reference Quaternion} 	\label{sec:App_orient}
	As presented in Section \ref{sec:RotationalMM}, the orientation of the 3D object is described by a reference quaternion, $\mathbf{q}_{\text{ref}}$, and a deviation vector, $\mathbf{a}$. 
	For maintaining this description, we follow the standard approach provided within the framework of MEKF, e.g., \cite{crassidis2007survey,markley2003attitude, maeder2011attitude}. 

	In particular, suppose that at time $k$, we have the reference quaternion $\mathbf{q}_{\text{ref}, k}$, and $\hat{\mathbf{a}}_{k|k}$ represents the estimated deviation vector, which is computed by the measurement update of the filter, given in Appendix \ref{sec:App_EKF_meas}. 
	We first update the reference quaternion in accordance with \eqref{eq:quatRef},		
	\begin{align}
		\mathbf{q}_{\text{ref}, k+1} = \delta\mathbf{q}(\hat{\mathbf{a}}_{k|k}) \odot \mathbf{q}_{\text{ref}, k}, 
	\end{align}
	where $\odot$ denotes the quaternion product defined in \eqref{eq:quatProd}. 
	Subsequently, the deviation vector is reset to zero as 
	\begin{align}
		\hat{\mathbf{a}}_{k|k} = \mathbf{0}. 
	\end{align}			
	
	Please note that the covariance of the deviation vector is kept unchanged, as suggested in \cite{markley2003attitude}. 
	This approach is referred to as the \textit{zero-order approximation}, \cite{gill2020full}, and it enables effective estimation of the orientation, as demonstrated by the comprehensive performance analysis. 
	For a detailed investigation of the reference orientation update mechanisms, interested readers can refer to \cite{gill2020full} and the references therein.			

	\ifCLASSOPTIONcaptionsoff
	\newpage
	\fi

	
	
	\bibliographystyle{IEEEtran}
	\bibliography{myBiblib}

\end{document}